\documentclass[a4paper,prd,11pt,noshowkeys,notitlepage,nofootinbib,superscriptaddress]{revtex4}
\usepackage[utf8]{inputenc}
\usepackage{ae,aecompl}
\usepackage{amsmath,amssymb,mathrsfs}
\usepackage{graphicx}
\usepackage{ccnishi}
\usepackage{dsfont}
\usepackage{slashed}
\usepackage{units}
\usepackage{mhchem}
\usepackage[normalem]{ulem}

\usepackage[usenames,dvipsnames]{xcolor} 
\usepackage{hyperref}
\hypersetup{
    colorlinks=true,        
    linkcolor=purple,       
    citecolor=JungleGreen,  
    filecolor=magenta,      
}

\providecommand{\lag}{\mathscr{L}} 
\providecommand{\Br}{\operatorname{Br}} 

\providecommand{\tPhi}{\tilde{\Phi}}

\providecommand{\la}[1]{\lambda^{(#1)}}
\providecommand{\muphi}{\mu_\varphi}
\newcommand{\amu}{\ensuremath{a_\mu}}

\providecommand{\Mphi}{M_{\varphi}}

\providecommand{\tf}{\tilde{f}}

\begin{document}
\title{
Connecting $(g-2)_\mu$ to neutrino mass in the extended neutrinophilic 2HDM
}
\author{A.~L.~Cherchiglia}
\email{alche@unicamp.br}
\affiliation{Instituto de F\'isica Gleb Wataghin, Universidade Estadual de Campinas, Campinas-SP, Brazil}
\affiliation{Departamento de F\'isica Te\'orica y del Cosmos, Universidad de Granada, Campus de Fuentenueva, E–18071 Granada, Spain}
\author{G.~De~Conto}
\email{george.de.conto@gmail.com}
\affiliation{Centro de Ci\^{e}ncias Naturais e Humanas,
Universidade Federal do ABC, Santo Andr\'{e}-SP, Brasil}
\author{C.~C.~Nishi}
\email{celso.nishi@ufabc.edu.br}
\affiliation{Centro de Matem\'{a}tica, Computa\c{c}\~{a}o e Cogni\c{c}\~{a}o,
Universidade Federal do ABC, Santo Andr\'{e}-SP, Brasil}

\begin{abstract}
One simple way to lower the scale of the seesaw mechanism that generates neutrino masses is to attribute part of their smallness to a suppressed vacuum expectation value of a second Higgs doublet as in the neutrinophilic 2HDM or in the type IB seesaw model.
On that structure we add one charged singlet scalar 
to induce a chirally enhanced contribution to $(g-2)_\mu$ with the same righthanded neutrinos of the seesaw.
We discuss the interplay of generating the necessary contribution to the latter with lepton flavor violation which is also necessarily brought to low scale.
We show that it is possible to explain $(g-2)_\mu$ even for heavy neutrino masses of order of a few TeV.

\end{abstract}
\maketitle
\section{Introduction}
\label{sec:intro}

The observation of neutrino oscillations firmly established that neutrinos have tiny masses and mix in the weak charged current\,\cite{ohlsson}. These properties clearly demonstrate that family lepton numbers associated to the lepton flavors $e,\mu,\tau$ are not conserved in nature.
The origin of both family lepton number violation and of neutrino masses can be naturally attributed to the dimension five Weinberg operator\,\cite{weinberg.op} pointing to a natural scale of $10^{12}\,\unit{GeV}$. 
If the new physics violating family lepton number appears only at this scale, then its effects, visible primarily on dimension six operators, are expected to be unobservably small.
These effects include those of charged lepton flavor violation (CLFV) that are expected to be probed with much greater precision in the coming years.

In contrast, the discrepancy between the experimental and SM prediction for the muon anomalous magnetic moment, $(g-2)_\mu$ or $\amu$, is a persistent anomaly that requires TeV scale (or lower) new physics coupling to the muon that may or may not violate lepton flavor.
A dedicated program to decrease the experimental uncertainty by a factor of four is currently ongoing in Fermilab, with the first data analysis released in 2021, and the second in 2023. 
The value for $\amu$ combined from the results obtained in Fermilab and Brookhaven is\;\cite{Muong-2:2021ojo}
$
	\amu^\text{Exp} = (11659205.9\pm2.2)\times10^{-10},
$
while the SM prediction from the Muon $g-2$ Theory Initiative White Paper\;\cite{Aoyama:2020ynm,g-2:theory} is
$
	\amu^{\text{SM}} = (11659181.0\pm4.3)\times10^{-10},
$
resulting in the $5.1\sigma$ discrepancy:
\begin{equation}
	\delta\amu^{\text{BSM}} = (24.9\pm4.8)\times10^{-10}.
        \label{eq:amuBSM}
\end{equation} 
The SM prediction makes use of dispersion methods to calculate the hadronic contributions that are the major source of uncertainty.\footnote{%
There is currently no consensus between the dispersion method and lattice results reported by the BMW collaboration\,\cite{BMW}. On the one hand, if the latter is used, the $\amu$ discrepancy is reduced to $1.5\sigma$ level. On the other hand, it disagrees with the R-ratio based prediction used in \cite{Aoyama:2020ynm} by more than $2.1\sigma$.
Comparison between different lattice groups is under way, with consistent results among them \cite{lattice}. 
Nevertheless, most of these results stand for specific euclidean windows, not the full determination of hadronic contributions to $\amu$. So an investigation of the source of incompatibility among the two approaches is urgent, together with more high precision lattice results. See \cite{Colangelo:2022vok} for an recent evaluation of the different euclidean windows using the dispersion method.
In this work, we will adopt the result in eq.\,\eqref{eq:amuBSM}.
}

By disregarding neutrino mass generation and minimally introducing new physics that couples to the muon in a flavor conserving fashion or obeying minimal flavor violation, one can explain $\amu$\,\cite{strumia,freitas,crivellin:g-2,calibbi.ziegler} and still easily avoid large CLFV effects.
A critical review of the minimal extensions with one or two new fields can be seen in Ref.\,\cite{stockinger};
see also Ref.\,\cite{queiroz}.
If the dominant new physics contribution occurs at 1-loop, a key ingredient is the presence of a \emph{chiral enhancement} that occurs by replacing one chiral flipping coupling involving the muon Yukawa $y_\mu\sim 0.0006$ by an order one coupling\,\cite{crivellin:g-2,calibbi.ziegler,curtin,strumia,yin.yamaguchi}.

With new physics at the TeV scale, the dipole operator responsible for $a_\mu$ should be given by the dimension 6 operator $\bar{\ell}_\mu\sigma^{\alpha\beta}H\mu_RF_{\alpha\beta}$ in the electroweak symmetric phase where the necessary chirality flip $\mu_R\to \mu_L$ involves a Higgs insertion.
Then the one-loop contribution to $\amu$ can be separated into two classes\,\cite{calibbi.ziegler,curtin} depending on the location of the Higgs insertion: (i) if it occurs in the external muon legs; (ii) if it occurs in one of the internal lines.
For contributions of type (i), there is no chiral enhancement because the Higgs insertion is accompanied by the usual muon Yukawa $y_\mu$ which also supplies a chiral flipping.\footnote{%
It is possible to explain $a_\mu$ with a type (i) contribution but either the new states in the loop need to lie at the electroweak scale or lower; see e.g. \cite{babu.lindner} or the couplings are at the edge of perturbativity; see e.g. \cite{renato}}.
If all new fields are electroweak singlets, this is the only possibility.
In contrast, chiral enhancement is possible for contributions of type (ii) because the Higgs is attached to a vertex other than the muon Yukawa.
Enhancement occurs if this vertex is replaced by the top Yukawa coupling (e.g., with the mediation of leptoquarks\,\cite{lepto.quark}) or the tau Yukawa coupling (e.g., with lepton flavor changing coupling with an additional Higgs doublet\,\cite{barman}).
The large coupling may also come from non-SM interactions such as with vectorlike leptons\,\cite{crivellin:g-2,dermisek}.
In all these examples, the new Higgs insertion also provides a chiral flipping.

With the presence of new scalars mixing with the SM Higgs, enhancement is also possible through the couplings of these new scalars with the muon.
In the Two-Higgs-doublet model (2HDM), this feature is available for the types II/X or aligned version with contributions enhanced by $\tan\beta$ (or, similarly, $\zeta_l$ for the aligned version). However, in order to accommodate $a_{\mu}$, some of the scalars must be at the weak scale, which renders only type X\,\cite{2hdm.typeX} or the aligned version\,\cite{Ilisie,Han,Cherchiglia1,Cherchiglia2} as possible candidates.
For the muon specific 2HDM\,\cite{mu-spec}, it is possible to push the scalars to higher masses.
In the MSSM, a similar reasoning apply, where the main contributions are also enhanced by $\tan\beta$, allowing to accommodate $a_\mu$ (see e.g. \cite{g-2:susy} for a review).
However, differently from the 2HDM versions mentioned, in the MSSM there are innumerous sources of flavor violation that need to be properly suppressed\,\cite{endo.yin,hisano}; see e.g. \cite{stockinger} for an updated analysis.

In this work, we aim to connect the physics responsible for $\amu$ with neutrino masses generation so that the same mediators participate in both processes.
We focus on TeV scale mediators and we also seek minimality in their number.
Two general features immediately follow.
Firstly, the seesaw scale has to be lowered to the TeV scale or lower and, secondly, one certainly cannot impose family lepton number to be conserved so that the interplay with CLFV is crucial. 
Simply lowering the mediator masses for the type I\,\cite{typeI}, type II\,\cite{typeII} or type III\,\cite{typeIII} seesaw models do not work because their contribution to $\amu$ is negative\,\cite{freitas,typeII:g-2}.\,\footnote{%
Ref.\,\cite{ma.raidal} had considered the type II seesaw and its interplay with $a_\mu$ but they did not consider the correct sign of the contribution.
}
In particular, adding any number of singlet RHNs to implement the type I seesaw mechanism leads to a negative contribution to $a_\mu$\,\cite{coy.frigerio} even in low-scale seesaw scenarios such as the inverse seesaw\,\cite{inverse.ss}
or the symmetry protected case\,\cite{kersten.smirnov}.
This situation does not change if the heavy particles do not participate in the generation of neutrino masses.

Focusing on the seesaw, there are many ways that the seesaw scale can be lowered. 
The aforementioned inverse seesaw is the case with extended tree level mediators that brings additional suppression due to approximate lepton number conservation. Another option is to consider radiative generation; see Ref.\,\cite{cai} for a review and Refs.\,\cite{radiative:g-2,barman,dicus.he} for models connecting $a_\mu$ with radiative neutrino mass generation. 
Yet another way is to attribute part of the smallness of neutrino masses to a vacuum expectation value (VEV) of a second Higgs doublet that only couples to neutrinos.
This is the neutrinophilic 2HDM ($\nu$-2HDM) for which the RHNs can be Majorana\,\cite{ma:nu2HDM,ma.raidal} or Dirac\,\cite{nandi,nu2hdm:U(1)}. 
Our focus will be on the former as the Majorana mass will induce the chirality flip enhancing the $a_\mu$ contribution.
An earlier attempt to connect neutrino masses with $a_\mu$ considered this neutrinophilic 2HDM with Majorana RHNs\,\cite{ma.raidal}, where the interplay with CLFV was studied.
However, in its simplest form, the new contribution to $(g-2)_\mu$ were negative.
Phenomenological studies unrelated to $a_\mu$ were performed in Refs.\,\cite{nu2HDM:pheno,haba.yamada}.
A second Higgs doublet with small vev also helps explaining the small neutrino mass in the recently proposed type IB seesaw\,\cite{fu.king} where two Higgs doublets couple to the two RHNs that form a pseudo-Dirac pair of an approximate $U(1)$ symmetry.
Other recent works connecting $a_\mu$ with neutrino mass generation can be seen in Ref.\,\cite{renato,nu.g-2:others}.

Here we show that the shortcomings of the $\nu$-2HDM (and also the type IB seesaw) to solve $a_\mu$ can be overcome by introducing another field ---a charged scalar singlet--- to the model so that a chirally enhanced contribution can be generated.
This contribution has no definite sign and allows the new fields to be at the TeV scale. Both the Majorana RHNs and the charged scalar contributes to $a_\mu$ while the RHNs and the neutrinophilic Higgs doublets are responsible for neutrino masses.
In particular, the RHN Majorana mass in the loop induces the necessarily chirality flip 
while the Higgs insertion occurs with scalars and it is decoupled from the chiral flipping.
Since lepton flavor violation is built in with the TeV scale new physics, it is important to study the induced CLFV processes.

The outline of the paper is as follows.
In Sec.\,\ref{sec:models} we present the models and in Sec.\,\ref{sec:nu.masses} we show how they generate the neutrino masses with TeV scale mediators.
Section \ref{sec:dipole} shows the calculation for $\amu$ and CLFV observables.
Section \ref{sec:solving} discusses the interplay between obtaining the correct $\amu$ and avoiding the constraints of CLFV processes.
The summary is given in Sec.\,\ref{sec:summary}.

\section{Models}
\label{sec:models}

\begin{figure}[h]
\includegraphics[scale=0.07]{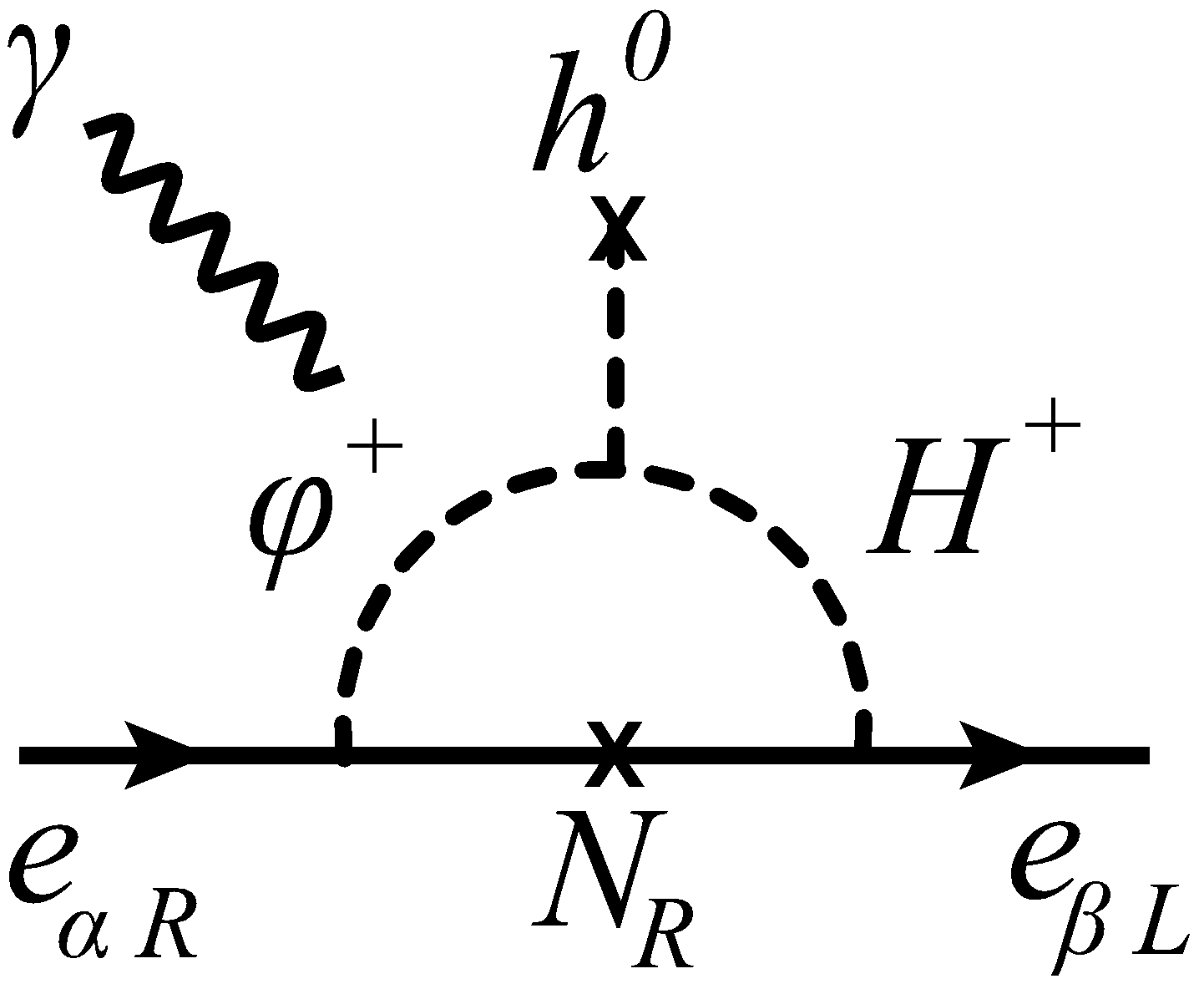}
\caption{\label{fig:g-2:NRphi}%
Chirally enhanced contribution to muon $g-2$ involving RHN $N_R$ and charged scalars.
}
\end{figure}

We seek scenarios where $a_\mu$ is explained with the participation of the righthand neutrinos (RHNs) at one-loop at the same time that these heavy neutrinos generate the necessary active neutrino masses through a low scale seesaw mechanism.
More specifically, we require a chiral enhanced contribution to $a_\mu$ connecting $\mu_L$ with $\mu_R$ and then two more charged scalars ---one electroweak singlet and another residing in a doublet--- are necessary to close the loop as shown in Fig.\,\ref{fig:g-2:NRphi}.
From the diagram it is clear that the chiral enhancement will be proportional to the RHN Majorana mass $M_N$ which we require to be at the TeV scale.\footnote{%
This is one of the minimal possibilities considered in Ref.\,\cite{calibbi.ziegler} to explain $(g-2)_\mu$ and DM stabilized by an unbroken $\ZZ_2$ leading to an inert Higgs doublet. Here the additional Higgs doublet has a small vev that helps explaining small neutrino masses.
}

So we extend the SM by considering two Higgs doublets $\Phi_1, \Phi_2$, two\,\footnote{%
These two neutrinos will generate the minimal number of two nonzero light neutrino masses but three RHNs can be equally considered.}
righthanded neutrino (RHN) fields $N_{iR}$, $i=1,2$, and one singlet charged scalar $\varphi^+$.
The relevant part of the Lagrangian involving the Higgs doublets and the righthanded neutrino fields is
\eq{
\label{lag:N}
-\lag\supset \bar{\ell}_\alpha h_\alpha \Phi_e e_{\alpha R}
+\bar{N}_{iR}\la{1}_{i\alpha}\tPhi_1^\dag \ell_\alpha
+\bar{N}_{iR}\la{2}_{i\alpha}\tPhi_2^\dag \ell_\alpha
+\ums{2}\bar{N}_{iR}(M_R)_{ij}N_{jR}^c+h.c.,
}
where $\ell_\alpha$, $\alpha=e,\mu,\tau$ are the lepton doublets and $\Phi_e$ is $\Phi_1$ or $\Phi_2$ depending on the model.
The relevant interaction terms involving the charged singlet scalar is 
\eq{
\label{lag:varphi}
-\lag\supset \muphi\Phi_2^\tp\epsilon\Phi_1\varphi^-+f_{i\alpha}\bar{N}_{iR}e^c_{\alpha R}\varphi^-
+h.c.,
}
where we choose $\muphi$ real without loss of generality.
The interaction term $\overline{\ell^c_\alpha}\ell_\beta\varphi^+$ is absent 
owing to the symmetries of the specific models introduced below and then there is no radiative neutrino mass generation as in the Zee model\,\cite{zee.model}.%
\footnote{A recent paper considered the Zee model for $(g-2)_{e,\mu}$\,\cite{barman}, where the couplings to charged singlet are not relevant and $a_\mu$ is explained by the contributions of the additional Higgs doublet with a chiral enhancement at one-loop of at most the $\tau$ Yukawa.
Ref.\,\cite{dicus.he} considered the complementary case of two charged scalar singlets and one RHN where neutrino masses are radiative but there is no chirally enhanced contribution to $a_\mu$. 
}
This absence also forbids the generation of four-fermion operators $\ell^4$ from tree level $\varphi^+$ exchange.

One of the simplest models within the type I seesaw is to suppress the Dirac mass term by attributing its origin to a Higgs doublet different from the rest, with tiny vacuum expectation value.
This is the neutrinophilic 2HDM model ($\nu$-2HDM)\,\cite{ma:nu2HDM} which is obtained from \eqref{lag:N} by imposing a $\ZZ_2$ where $N_{iR}, \Phi_1$ are odd. 
As a consequence, $\la{2}=0$ and $\Phi_e=\Phi_2$.
The singlet scalar $\varphi^+$ is additional and we assume it is odd so that 
the terms in \eqref{lag:varphi} are allowed. The resulting Lagrangian is 
\eqali{
\label{lag:nuphilic}
-\lag_{\nu\text{-2HDM}}&\supset \bar{\ell}_\alpha h_\alpha \Phi_2 e_{\alpha R}
+\bar{N}_{iR}\la{1}_{i\alpha}\tPhi_1^\dag\ell_\alpha
+\ums{2}\bar{N}_{iR}M_{N_i}N_{iR}^c
\cr 
&\quad +\ \muphi\Phi_2^\tp\epsilon\Phi_1\varphi^-+f_{i\alpha}\bar{N}_{iR}e^c_{\alpha R}\varphi^-
+h.c.
}
The doublet $\Phi_2\approx H_{\rm SM}$ will be mostly the SM Higgs doublet while $\Phi_1\approx H_\nu$ will be mostly composed of non-SM higgses. 
Neutrino masses will depend solely on $\aver{\Phi_1^0}=v_1$ which will be suppressed.

Another minimal possibility is to combine the righthanded neutrinos into a (pseudo)Dirac pair through a $U(1)_N$ symmetry with charges\,\cite{fu.king}
\eq{
N_{1R}\sim \Phi_1 \sim +1\,,\quad
N_{2R}\sim \Phi_2 \sim -1\,.
}
It leads to the Lagrangian
\eq{
\label{lag:IB:1}
-\lag_{\rm Ib}\supset \bar{N}_{R1}\lambda_{1\alpha}\tPhi_1^\dag\ell_\alpha
+\bar{N}_{R2}\lambda_{2\alpha}\tPhi_2^\dag\ell_\alpha
+M\overline{N_{R1}^c}N_{R2}
+h.c.
}
and the seesaw mechanism is dubbed seesaw type IB.
Compared to \eqref{lag:N}, we are already simplifying the notation in that $\la{1}_{1\alpha}=\lambda_{1\alpha}$ and $\la{2}_{2\alpha}=\lambda_{2\alpha}$ whereas 
$\la{1}_{2\alpha}=\la{2}_{1\alpha}=0$.
The other couplings depend on the choice of $U(1)_N$ charges for $e_{R\alpha}$ and $\varphi$.
We adopt the charges $e_R\sim -1$ and $\varphi^-\sim -2$ so that 
\eq{
\label{lag:IB:2}
-\lag_{\rm Ib}\supset 
h_\alpha\bar{\ell}_\alpha\Phi_1 e_{R\alpha}
+f_{2\alpha}\bar{N}_{2R}e^c_{\alpha R}\varphi^-
+\ \muphi\Phi_2^\tp\epsilon\Phi_1\varphi^-
+h.c.
}
To keep $h_\tau$ within perturbative values, we need $v_1\gtrsim 10^{-2}v$ and most of the suppression for light neutrino masses should still come from the suppression of the Yukawas $\lambda_{j\alpha}$.
We could flip the charges of $e_R$ and $\varphi^-$ and then replace $\Phi_1$ by $\Phi_2$ in the first term and $N_{2R}$ by $N_{1R}$ in the second.
But even in this case we cannot suppress $v_1$ too much because the dominant $g-2$ contribution is proportional to $v_1\lambda_{2\alpha}$.
The third term softly breaks the $U(1)_N$ symmetry\,\footnote{
The quadratic term $\Phi_2^\dag\Phi_1$ should be added as well.
The term $\ell\ell\varphi^+$ remains forbidden.
}
which cannot be exact as a global symmetry to avoid unwanted massless scalars.
The model in Ref.\,\cite[a]{fu.king}, for example, imposes a $\ZZ_3$ instead of $U(1)_N$.

For both models, the charged scalar component that is dominant in the non-SM Higgs doublet behaves as the charged scalar in the type I 2HDM and they are only constrained by LEP\,\cite{ALEPH:2013htx}:
\eq{
M_{H^+}>75\,\unit{GeV}\,.
}
On the other hand, the singlet component couples solely with charged leptons and righthand neutrino, implying that the model independent constraint from LEP does not apply. 
Since we allow for a small mixing, we will consider that both charged scalars will have masses around the electroweak scale or above.

\section{Light neutrino masses}
\label{sec:nu.masses}

Light neutrino masses are generated from the type I seesaw, i.e., through the exchange of RHNs in \eqref{lag:N} at tree level. The generated effective Weinberg operator is
\eq{
\label{weinberg.op}
\lag=\frac{1}{2}\bar{\ell}^c_{\alpha}\Gamma^\tp_{\alpha i}(M_R^{-1})_{ij}\Gamma_{j\beta}
\ell_\beta+h.c.\,,
}
depending on the two Higgs doublets through
\eq{
\label{Gamma.weinberg}
\Gamma_{j\beta}=\la{1}_{j\beta}\tPhi_1^\dag+\la{2}_{j\beta}\tPhi_2^\dag\,.
}

\subsection{$\nu$-2HDM model}

For the $\nu$-2HDM model, only $\Phi_1$ participates in the seesaw and $\la{2}=0$.
Then the neutrino mass matrix is given by
\eq{
\label{Mnu:nu2hdm}
M_\nu=-v_1^2{\la{1}}^\tp M_R^{-1}\la{1}\,,
}
where its lightness is partly explained by $\aver{\Phi_1^{0}}=v_1\ll v=\sqrt{v_1^2+v_2^2}=174\,\unit{GeV}$.
If $M_R\sim \unit{TeV}$, we need
\eq{
\label{v1:nu2hdm}
v_1\la{1}/v\sim \frac{\sqrt{M_R M_\nu}}{v}
\lesssim \frac{\sqrt{1\,\unit{TeV}\times\,0.05\,\unit{eV}}}{v}
\sim 10^{-6}\,.
}
We can choose $v_1/v\sim 10^{-6}$ in order to have $\la{1}$ of order one to address the deviation in $a_\mu$.

Since we are focusing on the minimal case of two RHNs, the heavy neutrino Yukawa coupling $\la{1}$ is mostly fixed by the masses and mixing of light neutrinos.
Using the Casas-Ibarra parametrization, we can write for normal ordering (NO),
\eqali{
\label{nu2HDM:lambda}
v_1\la{1}_{1\alpha} &= i\sqrt{M_1}\big(\sqrt{m_2}c_zV^\dag_{2\alpha}-\sqrt{m_3}s_zV^\dag_{3\alpha}\big)\,,
\cr
v_1\la{1}_{2\alpha} &= i\sqrt{M_2}\big(\sqrt{m_2}s_zV^\dag_{2\alpha}+\sqrt{m_3}c_zV^\dag_{3\alpha}\big)\,,
}
where $c_z=\cos z$ and $s_z=\sin z$ depend on the free complex angle $z$.
We are taking $M_R=\diag(M_1,M_2)$ and the neutrino mass matrix in \eqref{Mnu:nu2hdm} is $M_\nu=V^*\diag(0,m_2,m_3)V^\dag$, with $V$ being the PMNS matrix.
Therefore, besides the two CP phases in $V$, we have five free parameters in the neutrino sector: $M_1,M_2,\re(z),\im(z),\tan\beta=v_2/v_1$.
Note that for $M_1=M_2$, the real part of $z$ is not physical. 

For inverted ordering (IO), we have instead 
\eqali{
v_1\la{1}_{1\alpha} &= i\sqrt{M_1}\big(\sqrt{m_1}c_zV^\dag_{1\alpha}-\sqrt{m_2}s_zV^\dag_{2\alpha}\big)\,,
\cr
v_1\la{1}_{2\alpha} &= i\sqrt{M_2}\big(\sqrt{m_1}s_zV^\dag_{1\alpha}+\sqrt{m_2}c_zV^\dag_{2\alpha}\big)\,.
}
The only difference in this case is that $M_\nu=V^*\diag(m_1,m_2,0)V^\dag$.

\subsection{Type Ib seesaw model}

For the type IB seesaw model, the neutrino mass matrix coming from \eqref{weinberg.op} is
\eq{
(M_\nu)_{\alpha\beta}= -\frac{v_1v_2}{M}(\lambda_{1\alpha}\lambda_{2\beta}
    +\lambda_{2\alpha}\lambda_{1\beta})\,,
}
where $v_a=\aver{\Phi_a^0}$, $a=1,2$, and the Yukawa couplings are defined in \eqref{lag:IB:1}.
We choose $v_1<v_2$ because $\Phi_2$ is the doublet coupling to the charged leptons in \eqref{lag:IB:2}.
Adapting \eqref{v1:nu2hdm} to this case with $M\sim \unit{TeV}$, we obtain
\eq{
\label{v1v2:typeIb}
\frac{1}{v}\sqrt{v_1\lambda_{1\alpha}v_2\lambda_{2\beta}}\sim \frac{\sqrt{M_R M_\nu}}{v}
\lesssim \frac{\sqrt{1\,\unit{TeV}\times 0.05\,\unit{eV}}}{v}
\sim 10^{-6}\,.
}
As in the $\nu$-2HDM model, the Yukawa couplings $\lambda_{1\alpha}$ and $\lambda_{2\alpha}$ are mostly fixed by light neutrino masses and mixing.

For NO, the Yukawa couplings can be parametrized as
\eqali{
\label{typeIb:lambda:NO}
\lambda_{1\alpha}&=
i\kappa\left(\frac{M}{2v_1v_2}\right)^{1/2}
\bigg[
+ i\sqrt{m_2}\,V^\dag_{2\alpha}+\sqrt{m_3}\,V^\dag_{3\alpha}
\bigg]\,,
\cr
\lambda_{2\alpha}&=
i\kappa^{-1}\left(\frac{M}{2v_1v_2}\right)^{1/2}
\bigg[
- i\sqrt{m_2}\,V^\dag_{2\alpha}+\sqrt{m_3}\,V^\dag_{3\alpha}
\bigg]
\,,
}
where we use a different but equivalent parametrization with respect to Ref.\,\cite{fu.king}.
Flipping the sign in front of $\sqrt{m_2}$ is equivalent and this sign can be absorbed by flipping the sign of $\nu_{2L}$.
The parameter $\kappa$ is free and can be chosen real by rephasing $N_{1R}, N_{2R}$ with opposite phases.
For IO, we can analogously parametrize
\eqali{
\label{typeIb:lambda:IO}
\lambda_{1\alpha}&=
i\kappa\left(\frac{M}{2v_1v_2}\right)^{1/2}
\bigg[
+ i\sqrt{m_1}\,V^\dag_{1\alpha}+\sqrt{m_2}\,V^\dag_{2\alpha}
\bigg]\,,
\cr
\lambda_{2\alpha}&=
i\kappa^{-1}\left(\frac{M}{2v_1v_2}\right)^{1/2}
\bigg[
- i\sqrt{m_1}\,V^\dag_{1\alpha}+\sqrt{m_2}\,V^\dag_{2\alpha}
\bigg]
\,.
}
Note that in the neutrino sector, besides the unknown CP phases, there are only three free parameters: $M,\tan\beta, \kappa$. This is the same number as the neutrinophilic case with equal masses for the RHNs.

\section{Dipole moments and CLFV}
\label{sec:dipole}

\begin{figure}[h]
\includegraphics[scale=0.07]{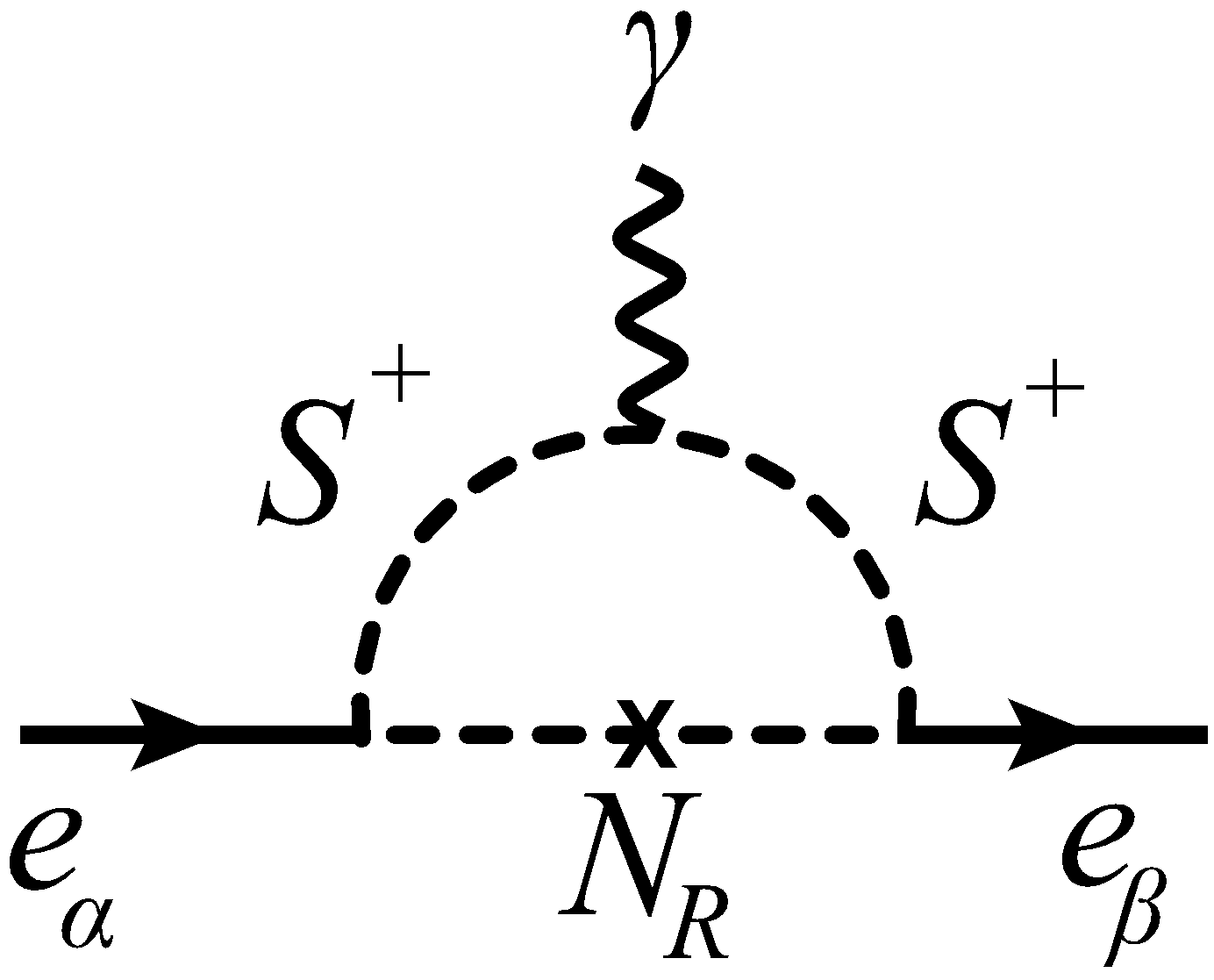}
\hspace{3em}
\includegraphics[scale=0.07]{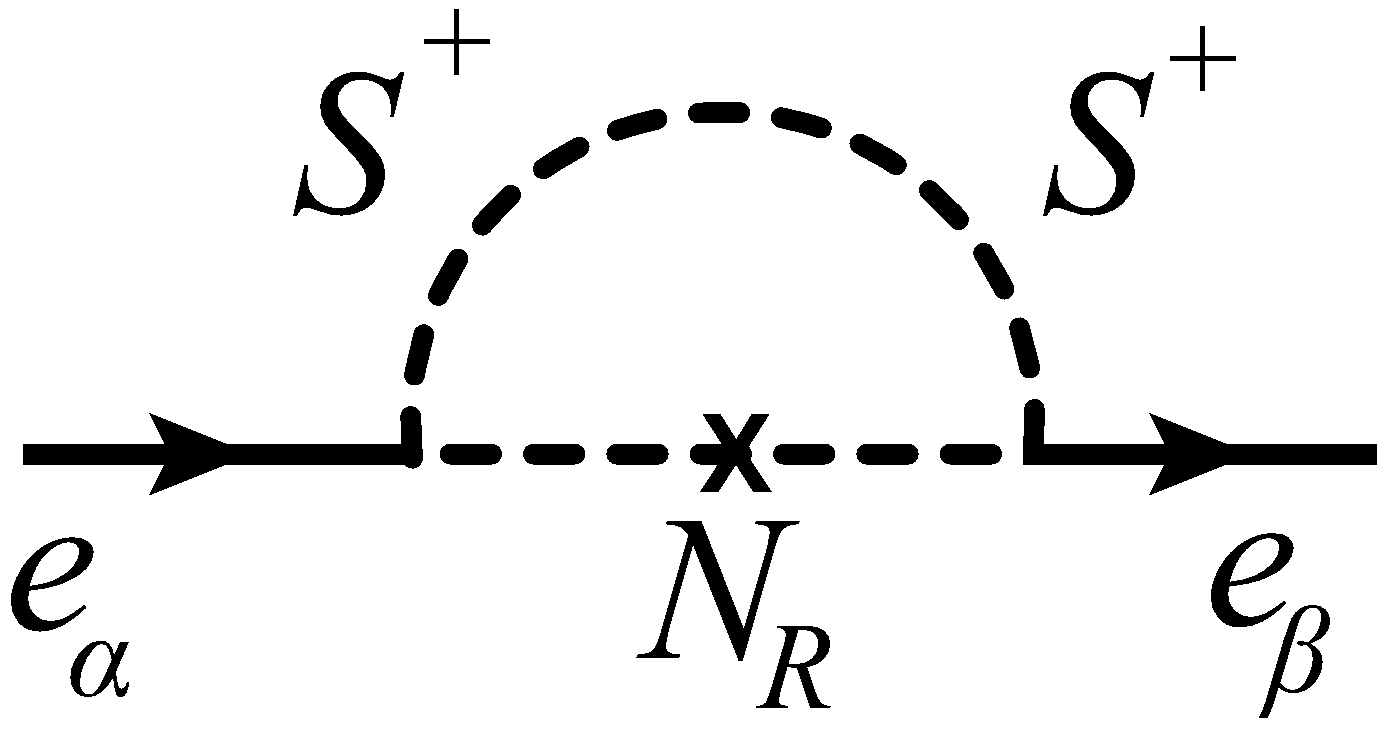}
\caption{\label{fig:dipole.self}%
Dipole and self-energy contribution to flavor changing processes.
}
\end{figure}

Using an effective theory approach, the operators relevant to lepton dipole moments and charged lepton flavor violation (CLFV) processes are the photonic operators
\eqali{
\label{L.eff:photon}
\lag_{\gamma\text{-eff}}&=
-\left(C^{\sigma R}_{\alpha\beta}\bar{e}_{\alpha L}\sigma_{\mu\nu}e_{\beta R} F^{\mu\nu}
+h.c.\right)
\cr&\quad
-\ \left(C_{\alpha\beta}^{\text{ND-}L}\bar{e}_{\alpha L}\gamma_\nu e_{\beta L}
+C_{\alpha\beta}^{\text{ND-}R}\bar{e}_{\alpha R}\gamma_\nu e_{\beta R}
\right)\partial_\mu F^{\mu\nu}
\,.
}
The operator in the first line is the dipole contribution whereas the ones in the second line are the non-dipole (ND) part.
We can see that only the dipole part involves chirarity flipping and chiral enhancement will be possible only for this term.
The Wilson coefficients at 1-loop can be obtained by matching the full theory with the effective theory through appropriate 1-loop amplitudes.
The relevant amplitudes come from the dipole (left) and self-energy (right) diagrams in Fig.\,\ref{fig:dipole.self}.

\subsection{Dipole moments and $\ell_\alpha\to \ell_\beta\gamma$}

The Wilson coefficient of the dipole operator
contributes to the dipole moments and $\ell_\alpha\to \ell_\beta\gamma$ as
\eqali{
\label{a.mu:C}
a_{\alpha}&=-\frac{4 m_\alpha}{e}\re(C^{\sigma R}_{\alpha\alpha})\,,
\cr
d_{\alpha}&=-2\im(C^{\sigma R}_{\alpha\alpha})\,,
\cr
\Br[\ell_\alpha\to \ell_\beta\gamma] &=\frac{m^3_\alpha}{4\pi \Gamma_\alpha}(|C^{\sigma R}_{\alpha\beta}|^2+|C^{\sigma R}_{\beta\alpha}|^2)\,,
}
where $a_\alpha,d_\alpha$ are the contributions to the magnetic and eletric dipole moments, respectively.
These formulas assume that our covariant derivative in QED is $D_\mu=\partial_\mu+ieQA_\mu$ and similarly for the SM.

For the $\nu$-2HDM described in \eqref{lag:nuphilic}, $\Phi_2\approx H_{\rm SM}$ and $\Phi_1\approx H_\nu$ is almost inert with a tiny vev.
The chirally enhanced contribution to the dipole operator is given in Fig.\,\ref{fig:g-2:NRphi}, with $\varphi$ being the charged singlet, $H^+$ being essentially the charged Higgs residing in $\Phi_1$
and $h^0$ being the SM Higgs boson.
The chiral enhancement comes from the $N_{iR}$ Majorana masses, $M_{N_i}$, and these heavy neutrinos only couple to $\Phi_1$.
The expected chirally enhanced contribution is then of the form
\eq{
\label{Csigma.expected}
C^{\sigma R}_{\beta\alpha}\sim
\frac{ev_2\muphi}{(4\pi)^2\Lambda^4}{\la{1}_{\beta i}}^\dag M_{N_i}f^*_{i\alpha}\,,
}
where the transferred momentum is $q^2=-m_\mu^2$.

For the exact contribution we need to include the mixing between the charged higgs $H^+$ in the doublet and the scalar singlet $\varphi^+$ coming from the $\muphi$ term in \eqref{lag:nuphilic} or \eqref{lag:varphi}. The contribution in the scalar potential is
\eq{
\label{phi:mixing.term}
\muphi\varphi^-(\Phi_2^+\Phi_1^0-\Phi_2^0\Phi_1^+)\to
-\muphi v\,\varphi^-H^+\,,
}
where $vH^+\equiv v_2\Phi_1^+ -v_1\Phi_2^+$, $v_i=\aver{\Phi_i^0}$, is the physical charged Higgs field within the doublets when $\muphi=0$. The orthogonal direction is the charged Goldstone absorbed by the $W$.
Considering arbitrary mass terms for $H^+H^-$ and $\varphi^+\varphi^-$, the mixing term \eqref{phi:mixing.term} induces the mixing
\eq{
\mtrx{H^+\cr \varphi^+}=
\mtrx{c_\gamma & -s_\gamma\cr s_\gamma & c_\gamma}\mtrx{S_1^+\cr S_2^+}\,,
}
with angle
\eq{
\label{def.gamma}
\sin2\gamma= -\frac{2\muphi v}{M^2_{S_2}-M^2_{S_1}}\,,
}
and $S_i^+$ are the charged scalars with masses $M_{S_i}$, $i=1,2$.
We have chosen $M_{S_2}>M_{S_1}$.

Then the complete contribution arising from the diagrams in Fig.\,\ref{fig:dipole.self} is
\subeqali[C.dipole]{
\label{cR:lambda.f}
\frac{16\pi^2}{e}C^{\sigma R}_{\beta\alpha}
&=
c_\gamma s_\gamma \frac{v_2}{v}\sum_j\frac{{\lambda^{(1)}_{\beta j}}^\dag f^*_{j\alpha}}{M_{N_j}}\left[x_{2j}f_S (x_{2j})-x_{1j}f_S (x_{1j})\right]\,,
\\
\label{cR:lambda.lambda}
&\quad +\ 
m_\alpha \frac{v_2^2}{v^2}\sum_j\lambda^{(1)\dag}_{\beta j}
    \left[
    \frac{c_\gamma^2}{M^2_{S_1}}\tf_S(x_{1j})
    +\frac{s_\gamma^2}{M^2_{S_2}}\tf_S(x_{2j})
    \right]
\la{1}_{j\alpha}
\\
\label{cR:f.f}
&~~\ 
+
m_\beta \sum_j f^\tp_{\beta j}
    \left[
    \frac{s_\gamma^2}{M^2_{S_1}}\tf_S(x_{1j})
    +\frac{c_\gamma^2}{M^2_{S_2}}\tf_S(x_{2j})
    \right]
f_{j\alpha}^*
\,,
}
where $x_{kj}\equiv M^2_{N_j}/M^2_{S_k}$ and the loop functions are\,\cite{crivellin:g-2,calibbi.ziegler}
\eqali{
f_S (x)&\equiv \frac{x^2-1-2x \log x}{4(x-1)^3}\,,
\cr
\tf_S(x)&\equiv \frac{2x^3+3x^2-6x+1-6x^2\log x}{24(x-1)^4}\,.
}
See appendix \ref{ap:wilson} for the details.
We have checked that the contribution proportional to $(\la{1})^2$ matches Ref.\,\cite{haba.yamada} for $\gamma=0$.

Let us discuss the various contributions.
The first contribution \eqref{cR:lambda.f} is the chirally enhanced (left-right), reducing to \eqref{Csigma.expected} when $M_{S_i}\gg M_{N_j}$ as $f_S(0)=1/4$ and we identify $\Lambda^4=4M_{S_2}^2M_{S_1}^2(M_{S_2}^2-M_{S_1}^2)$.
The contributions \eqref{cR:lambda.lambda} and \eqref{cR:f.f} are not chirally enhanced (left-left and right-right respectively) as the chiral flip comes from the external lines.
So the chirally enhanced contribution is larger than the non-chirally enhanced contributions by a factor
\eq{
c_\gamma s_\gamma \frac{M^2_S}{m_\mu M_N}\sim c_\gamma s_\gamma\times 10^4\,,
}
for order one couplings and the numbers assume $M_N\sim M_{S_i}\sim \unit{TeV}$.
Given the definition of the angle $\gamma$ in \eqref{def.gamma} the trilinear coupling $\mu_\varphi$ cannot be arbitrarily small if we require the chirally enhanced contribution to dominate:
\eq{
\label{muphi>}
|\mu_\varphi|\gtrsim 1\,\unit{GeV}
\text{~~or~~}
|s_\gamma|\gtrsim 10^{-4}
\,.
}
The dominance of the chirally enhanced contribution is important because the non-chirally enhanced contributions are positive definite and leads to a contribution to $a_\mu$ which is negative definite, contrary to the experimental observation.
They are similar to the contribution of $N_R$ exchange in the usual seesaw models\,\cite{freitas,coy.frigerio}.

For the type IB model in \eqref{lag:IB:1} and \eqref{lag:IB:2}, the estimate of the chirally enhanced contribution analogous to \eqref{Csigma.expected}  will be
\eq{
\label{Csigma.expected:Ib}
C^{\sigma R}_{\beta\alpha}\sim
\frac{v_2\muphi}{(4\pi)^2\Lambda^4}{\lambda}^\dag_{\beta 1} Mf^*_{2\alpha}\,.
}
Then we need $\lambda_{1\beta}$ to be of order one while $\lambda_{2\beta}$ needs to be suppressed 
due to neutrino masses, cf.\,\eqref{v1v2:typeIb}.

The complete contribution reads
\subeqali[C.dipole:Ib]{
\label{cR:lambda.f:Ib}
\frac{16\pi^2}{e}C^{\sigma R}_{\beta\alpha}
&=
c_\gamma s_\gamma \frac{v_2}{v}\frac{\lambda^\dag_{\beta 1} f^*_{2\alpha}}{M_{N}}\left[x_{2}f_S (x_{2})-x_{1}f_S (x_{1})\right]\,,
\\
\label{cR:lambda.lambda:Ib}
&\quad +\ 
m_\alpha \frac{v_2^2}{v^2}\lambda^{\dag}_{\beta 1}
    \left[
    \frac{c_\gamma^2}{M^2_{S_1}}\tf_S(x_{1})
    +\frac{s_\gamma^2}{M^2_{S_2}}\tf_S(x_{2})
    \right]
\lambda_{1\alpha}
\\
\label{cR:f.f:Ib}
&~~\ 
+
m_\beta f^\tp_{\beta 2}
    \left[
    \frac{s_\gamma^2}{M^2_{S_1}}\tf_S(x_{1})
    +\frac{c_\gamma^2}{M^2_{S_2}}\tf_S(x_{2})
    \right]
f_{2\alpha}^*
\,,
}
where $x_i=M^2_N/M_{S_i^2}$. We have neglected additional contributions proportional to 
$\lambda_{2\beta}^*\lambda_{2\alpha}v_1^2$ which will be highly suppressed.
Here the requirement \eqref{muphi>} is equally necessary for order one couplings.

In Table \ref{tab:LFV} we show the current and future limits for different $\ell_\alpha\to \ell_\beta\gamma$.
Limits for other CLFV processes are also shown.
\begin{table}[h]
\[
\begin{array}{|c|c|c|c|}
\hline  \text{Observable}    & \text{Current limit} & \text{Future limit} 
    \cr
\hline
\text{Br}(\mu\to eee)       &      <1.0 \times 10^{-12} \text{\cite{PDG2020}}       &  10^{-16} \text{\cite{mu3e}} \cr
\text{Br}(\tau\to \mu\mu\mu)       &   <2.1 \times 10^{-8} \text{\cite{HFLAV:2019otj}}    & 3.4\times 10^{-10} \text{\cite{belle2:book}} \cr
\text{Br}(\tau\to \mu ee)       &  <8.4 \times 10^{-9} \text{\cite{HFLAV:2019otj}}  & 2.9\times 10^{-10} \text{\cite{belle2:book}}\cr    
\text{Br}(\tau\to eee)       &  <1.4 \times 10^{-8} \text{\cite{HFLAV:2019otj}}  & 4.3\times 10^{-10} \text{\cite{belle2:book}}\cr
\text{Br}(\tau\to e\mu\mu)       &  <1.6 \times 10^{-8} \text{\cite{HFLAV:2019otj}} & 4.3\times 10^{-10} \text{\cite{belle2:book}}\cr
\hline
\text{Br}(\mu\to e\gamma)       &  <4.2 \times 10^{-13} \text{\cite{PDG2020}} & 6\times 10^{-14}\text{\cite{meg2}} \cr
\text{Br}(\tau\to \mu\gamma)       &  <4.4 \times 10^{-8} \text{\cite{PDG2020}} & 10^{-9} \text{\cite{belle2:book}} \cr
\text{Br}(\tau\to e\gamma)       &  <3.3 \times 10^{-8} \text{\cite{PDG2020}} & 3\times 10^{-9} \text{\cite{belle2:book}} \cr 
\hline
\Gamma^{\rm conv}_{\mu \rightarrow e}/\Gamma^{\rm capt}_{N} & <7.0 \times 10^{-13} \text{\cite{SINDRUMII:2006dvw}}^{*} & 3 \times 10^{-17} \text{\cite{mu2e,comet}}^{**}/ 10^{-18}\text{\cite{jparc}}^{\dagger}\cr
\hline
\end{array}
\]
\caption{\label{tab:LFV}
Current and future limits for charged lepton flavor violating processes at 90\% CL.
For $\mu e$ conversion, the nucleous is Au (*) and Al (**) or Ti ($\dagger$). 
}
\end{table}

\subsection{$\mu$-$e$ conversion in nuclei}

A very stringent test for CFLV is the coherent $\mu^-\text{\,--\,}e^-$ conversion in a muonic atom of nucleus $(A,Z)$ by neutrinoless muon capture
\[
\mu^-+(A,Z)\to e^-+(A,Z)\,,
\]
mediated in our case by the effective flavor-changing photon interactions in \eqref{L.eff:photon}.
We neglect similar $Z$ mediated processes which are suppressed by the $Z$ mass. It should be noticed that, contrary to $\mu \rightarrow e \gamma$ decay, one needs to consider off-shell photon emission. The general photonic $\mu{-}e$ transition amplitude is given by \cite{Kuno}
\eq{
\mathcal{M} = -e A_{\mu}^{*}(q)\bar{u}_e(p_{e})\left[(f_{E0} + \gamma_{5} f_{M0})\gamma_{\nu}\left(g^{\mu\nu}-\frac{q^{\mu}q^{\nu}}{q^{2}}\right) + (f_{M1} + \gamma_{5} f_{E1})\frac{i\sigma_{\mu\nu}q^{\nu}}{m_{\mu}}\right]u_{\mu}(p_{\mu})\,,
}
where $p_{\mu}$ and $p_{e}$ are the momenta of the muon and electron respectively, $q=p_{\mu}-p_{e}$ is the transfered momentum, and $f_{X}\equiv f_{X}(q^{2})$ are form factors. In the case of $\mu\rightarrow e \gamma$ decay, only the dipole part ($f_{E1}$, $f_{M1}$) contributes. The form factors can be written in terms of the Wilson coefficients introduced in \eqref{L.eff:photon}:
\begin{align}
f_{E0}&=(C_{e\mu}^{\text{ND-}L}+C_{e\mu}^{\text{ND-}R})\frac{q^{2}}{2e}, \quad f_{M0}=(C_{e\mu}^{\text{ND-}L}-C_{e\mu}^{\text{ND-}R})\frac{q^{2}}{2e},\\
f_{E1}&=-(C_{e\mu}^{\sigma R}-C_{\mu e}^{\sigma R *})\frac{m_{\mu}}{e}, \quad f_{M1}=-(C_{e\mu}^{\sigma R}+C_{\mu e}^{\sigma R *})\frac{m_{\mu}}{e}\,.
\end{align}

In the approximation pioneered by Weinberg and Feinberg \cite{Weinberg:1959zz}, only the photonic contribution is included in the calculation of the branching ratio of the coherent $\mu-e$ conversion which is given by
\eq{
\Br[\mu N \to e N]\equiv\frac{\Gamma[\mu N \to e N]}{\Gamma_{\rm capt}} =\frac{8 m_{\mu} \alpha^5 Z_{\rm eff}^4 Z |F_{p}|^2\xi^{2}}{\Gamma_{\rm capt}}\,.
}
Here $Z_{\rm eff}$ is an effective atomic charge due to averaging the muon wave function over the nuclear density, $\Gamma_{\rm capt}$ is the total muon capture rate, and $\xi^{2}$ is obtained with knowledge of the form factors
\eq{
\xi^{2} =|f_{E0}(-m_{\mu}^{2})+f_{M1}(-m_{\mu}^{2})|^{2} + |f_{E1}(-m_{\mu}^{2})+f_{M0}(-m_{\mu}^{2})|^{2}\,,
}
where we use the transferred momentum $q^2=-m_\mu^2$.

In general, the non-photonic part should be included as well, which is parametrized by an effective four-fermion interaction containing quarks. In our particular case, these contributions would come from box diagrams containing both the righthanded neutrinos and charged scalars. Since the coupling between quarks and the charged scalars is suppressed (it will be proportional to the ratio $v_{1}/v_{2}$), we can safely neglect these diagrams.

Finally, we rewrite the branching ratio in terms of the Wilson coefficients, to allow a straightforward application to our model
\eq{
\label{mueconv:final}
\Br[\mu N \to e N] =\frac{4m^3_\mu \alpha^4 Z_{\rm eff}^4 Z |F_{p}|^2}{\pi \Gamma_{\rm capt}}\left(\left| C^{\sigma R}_{e\mu}+\frac{m_\mu}{2}C^{\rm ND-L}_{e\mu}\right|^2+\left|C^{\sigma R}_{\mu e}+\frac{m_\mu}{2}(C^{\rm ND-R}_{e\mu})^*\right|^2\right)\,.
}
In our case, for order one couplings and loop functions, the chiral enhanced part in the dipole contribution will dominate the non-dipole part by a factor $M_{S_i}/m_\mu\sim M_{N_k}/m_\mu\sim 10^4$ for the new fields in the TeV scale and $s_\gamma$ not too small as \eqref{muphi>}.
In this case of dipole dominance, the ratio $\Br[\mu N \to e N]/\Br(\mu\to e\gamma)$ depends roughly only on the atomic number of the nucleous\,\cite{cirigliano} and we have checked it for our case.
As for the non-dipole part, we have found that the relative sign between $C^{\sigma R}$ and $C^{\rm ND}$ in \eqref{mueconv:final} is effectively opposite to Refs.\,\cite{haba.yamada,bertuzzo.perez,arganda}.

At present, the best limit for this branching ratio comes from conversion in gold nuclei \cite{SINDRUMII:2006dvw}: $\Gamma_{\rm conv}^{\rm Au}/\Gamma_{\rm capt}^{\rm Au} < 7.0 \times 10^{-13}$, where $\Gamma_{\rm capt}^{\rm Au}= 8.7 \times 10^{-18}$ GeV. There are, however, future experiments that aim to reduce the bounds on $\mu-e$ conversion by several orders of magnitude \cite{mu2e,comet}. Their aim is to achieve $\Gamma_{\rm conv}^{\rm Al}/\Gamma_{\rm capt}^{\rm Al} < 3 \times 10^{-17}$ using aluminium nuclei.
The different effective parameters for different nucleous are shown in Table~\ref{tab:Zeff}.
\begin{table}[h]
\[
\begin{array}{|c|c|c|c|}
\hline  \ce{^{$A$}_{$Z$}Nucleus}    & Z_{\rm eff} & F_p & \Gamma_{\rm capt}(\unit{GeV}) 
    \cr
\hline
\ce{^{27}_{13}Al}    &  11.5  & 0.64 & 4.64 \times 10^{-19} \cr
\ce{^{48}_{22}Ti}    &  17.6  & 0.54 & 1.70 \times 10^{-18}       \cr
\ce{^{80}_{38}Sr}    &  25.0  & 0.39 & 4.62 \times 10^{-18}       \cr
\ce{^{121}_{51}Sb}    &  29.0  & 0.32 & 6.72 \times 10^{-18}       \cr
\ce{^{197}_{79}Au}    &  33.5  & 0.16 & 8.60 \times 10^{-18}       \cr
\ce{^{207}_{82}Pb}    &   \;34.0 \; &  \;0.15 \; & 8.85 \times 10^{-18}       \cr
\hline
\end{array}
\]
\caption{\label{tab:Zeff}%
Values for $Z_{\rm eff}$, $F_p$, and $\Gamma_{\rm capt}(\unit{GeV})$ for different nuclei\;\cite{Kitano,arganda}.}
\end{table}

The Wilson coefficient of the non-dipole part can be obtained for the $\nu$-2HDM model as
\eqali{
\label{CND:nu}
\frac{16\pi^2}{e}C_{\beta\alpha}^{\text{ND-}L}&=
\frac{v_2^2}{v^2}\sum_j{\la{1}_{\beta j}}^\dag\la{1}_{j\alpha}\left[
c_\gamma^2\frac{G_S(x_{1j})}{6M^2_{S_1}}+s_\gamma^2\frac{G_S(x_{2j})}{6M^2_{S_2}}
\right]\,,
\cr
\frac{16\pi^2}{e}C_{\beta\alpha}^{\text{ND-}R}&=
\sum_j f^\tp_{\beta j} f^*_{j\alpha}\left[
s_\gamma^2\frac{G_S(x_{1j})}{6M^2_{S_1}}+c_\gamma^2\frac{G_S(x_{2j})}{6M^2_{S_2}}
\right]\,,
}
where the loop function is
\eq{
\label{GS}
G_S(x)=\frac{2-9 x+18 x^2-11 x^3+6 x^3 \log (x)}{6 (1-x)^4}\,.
}
The latter function is the same as $G_2(x)$ in Ref.\,\cite{haba.yamada}.
See details in appendix~\ref{ap:wilson}.
As anticipated, there is no chiral enhancement for these coefficients.

For the type IB seesaw model, we analogously obtain
\eqali{
\label{CND:Ib}
\frac{16\pi^2}{e}C_{\beta\alpha}^{\text{ND-}L}&=
\frac{v_2^2}{v^2}\lambda^\dag_{\beta 1}\lambda_{1\alpha}\left[
c_\gamma^2\frac{G_S(x_{1})}{6M^2_{S_1}}+s_\gamma^2\frac{G_S(x_{2})}{6M^2_{S_2}}
\right]\,,
\cr
\frac{16\pi^2}{e}C_{\beta\alpha}^{\text{ND-}R}&=
f^\tp_{\beta 2} f^*_{2\alpha}\left[
s_\gamma^2\frac{G_S(x_{1j})}{6M^2_{S_1}}+c_\gamma^2\frac{G_S(x_{2j})}{6M^2_{S_2}}
\right]\,.
}

We could also consider the decay $\mu\rightarrow eee$. In addition to the Wilson coefficients that we already have, we will need to consider box diagrams. These diagrams have righthanded neutrinos and charged scalars inside, whose contributions will be of the form
\eq{
f_{i\mu}f_{ie}f_{je}f_{je},\quad \lambda_{i\mu}\lambda_{ie}\lambda_{je}\lambda_{je}, \quad \lambda_{i\mu}f_{ie}\lambda_{je}f_{je}, \quad f_{i\mu}\lambda_{ie}f_{je}\lambda_{je}\nonumber
}
In Sec.\,\ref{sec:solving} we will see that we need suppressed $f_{ie}$ to evade CLFV constraints and still be able to account for $(g-2)_\mu$.
In this case, only the contribution with $\lambda$ survives which is subdominant\,\cite{haba.yamada}.

\section{Solving $(g-2)_\mu$ avoiding CLFV}
\label{sec:solving}

By considering only the dominant chirally enhanced contribution to the dipole term, we can estimate the interplay between the necessary contribution to $a_\mu$ and the necessary suppression to avoid 
significant CLFV processes.
The contribution to $a_\mu$ in \eqref{a.mu:C} necessary to explain the experimental deviation \eqref{eq:amuBSM} requires
\eq{
|C^{\sigma R}_{\mu\mu}|\sim |\re C^{\sigma R}_{\mu\mu}|
\sim 2\times 10^{-9}\,\unit{GeV}^{-1}\,.
}
On the other hand, the current limit on $\mu\to e\gamma$ requires
\eq{
\sqrt{|C^{\sigma R}_{\mu e}|^2+|C^{\sigma R}_{e\mu}|^2}
<4\times 10^{-14}\,\unit{GeV}^{-1}\,.
}
So it is necessary that
\eq{
\label{ratio:muegamma/amu}
\frac{\sqrt{|C^{\sigma R}_{\mu e}|^2+|C^{\sigma R}_{e\mu}|^2}}{|C^{\sigma R}_{\mu \mu}|}
\lesssim 2\times 10^{-5}\,.
}
The expected future limit will decrease this number by a factor $2.6$.
A similar analysis on the current limit on $\mu e$ conversion in gold nuclei requires
\eq{
\label{mue:Csigma:current}
\sqrt{|C^{\sigma R}_{\mu e}|^2+|C^{\sigma R}_{e\mu}|^2}
< 10^{-12}\,\unit{GeV}^{-1}\,,
}
which implies
\eq{
\label{mue:ratio:current}
\frac{\sqrt{|C^{\sigma R}_{\mu e}|^2+|C^{\sigma R}_{e\mu}|^2}}{|C^{\sigma R}_{\mu \mu}|}
\lesssim 5\times 10^{-4}\,.
}
The expected future limit in aluminium nuclei will greatly reduce the limit in \eqref{mue:Csigma:current} to $6\times 10^{-15}\,\unit{GeV}^{-1}$ and then the bound in \eqref{mue:ratio:current} becomes $3\times 10^{-6}$.

Now, let us see how our models can satisfy the hierarchy \eqref{ratio:muegamma/amu} between flavor changing and flavor conserving couplings to the muon.

We start with the type IB seesaw model.
We assume the dominance of the chirally enhanced contribution and require $|s_\gamma|\gg 10^{-4}$ due to \eqref{muphi>}.
In this model, to have appreciable $a_\mu$, we need
\eq{
\lambda^*_{1\mu}f^*_{2\mu}\sim O(1)\,.
}
On the other hand, according to \eqref{ratio:muegamma/amu}, we need $|\lambda^*_{1\mu}f^*_{2 e}|, |\lambda^*_{1e}f^*_{2 \mu}|$ to be less than $10^{-5}$.
Since $\lambda_{1\mu}$ needs to be order one, we can adopt $f_{2e}=0$ for simplicity.
In this case,
\eq{
\frac{\sqrt{|C^{\sigma R}_{\mu e}|^2+|C^{\sigma R}_{e\mu}|^2}}{|C^{\sigma R}_{\mu \mu}|^2}
\approx\frac{|\lambda_{1e}|}{|\lambda_{1\mu}|}
\gtrsim 0.1\,,
}
from the dependence of $\lambda_{1\alpha}$ on neutrino masses and mixing for NO or IO, cf.\ eqs.\,\eqref{typeIb:lambda:NO} or \eqref{typeIb:lambda:IO}, and \eqref{ratio:muegamma/amu} is never satisfied.
Turning on the coupling $f_{2e}$ only worsens the situation and the type IB seesaw model augmented with a singly charged singlet cannot account for $(g-2)_\mu$ without violating current bounds on $\mu\to e\gamma$.

We now turn to the $\nu$-2HDM model.
For appreciable $a_\mu$, assuming the dominance of chirally enhanced contribution, we need large $\mu\mu$ couplings:
\eq{
\label{nu2HDM:O(1)}
{\la{1}_{\mu 1}}^\dag f^*_{1\mu},\,{\la{1}_{\mu 2}}^\dag f^*_{2\mu}\sim O(1)\,.
}
For equal masses $M_{1}=M_{2}$,
to suppress CLFV processes, we need suppressed $\mu e$ couplings:
\eq{
|{\la{1}_{\mu 1}}^\dag f^*_{1e}+{\la{1}_{\mu 2}}^\dag f^*_{2e}|\lesssim 10^{-5}\,,\quad
|{\la{1}_{e 1}}^\dag f^*_{1\mu}+{\la{1}_{e 2}}^\dag f^*_{2\mu}|\lesssim 10^{-5}\,.
}
Considering \eqref{nu2HDM:O(1)}, the first combination vanishes if 
\eq{
\label{fje=0}
f_{1e}=f_{2e}=0\,,
}
while the vanishing of the second requires the orthogonality between $f_{i\mu}^*$ and $\la{1}_{ie}$:
\eq{
\label{align.f}
(f_{1\mu},f_{2\mu})=\zeta (\la1_{2e},-\la1_{1e})\,.
}
The contribution to $\amu$ then is proportional to
\eq{
\label{lambda.f:align}
{\la{1}_{\mu 1}}^\dag f^*_{1\mu}+{\la{1}_{\mu 2}}^\dag f^*_{2\mu}
=\zeta({\la1_{2e}}^*{\la1_{1\mu}}^*-{\la1_{1e}}^*{\la1_{2\mu}}^*)\,,
}
which is mostly fixed from neutrino parameters.
The mass degeneracy $M_{1}=M_{2}$ and the orthogonality condition \eqref{align.f} could in principle be justified by flavor symmetries\,\cite{feruglio} which needs much more structure and will not be treated here. 
This links the coupling $f$ with $\lambda$ so that all the terms in \eqref{C.dipole} scale as $1/v_1^2$.
The vanishing \eqref{fje=0} also makes the BSM contribution to electron EDM negligible, making it easily compatible with the current precise measurement\,\cite{d.e}.
For the muon EDM, the combination \eqref{lambda.f:align} contributes but it is equally negligible compared to the current limit\,\cite{d.mu}.

Considering now the complete contribution to the dipole coefficient $C^{\sigma R}_{\beta\alpha}$ in \eqref{C.dipole}, the turning off (or suppression) of the electron coupling \eqref{fje=0} eliminates the $f^2$ contribution \eqref{cR:f.f} for $\alpha=e$ or $\beta=e$.
Analogously, the orthogonality condition \eqref{align.f} eliminates the mixed (left-right) contribution \eqref{cR:lambda.f}, i.e., the chiral enhanced contribution, to the flavor transition $\mu\to e$ if $N_1,N_2$ have equal masses.
Then, only the $\lambda^2$ term (left-left) in \eqref{cR:f.f}, which is not enhanced, contributes to $\mu\to e \gamma$ while the chiral enhanced part still contributes dominantly to $a_\mu$.
With the increasing of the mass difference $\Delta M_N=M_2-M_1$, the chiral enhanced contribution also increases rapidly for $\mu\to e\gamma$.
In principle, we could change the orthogonality condition \eqref{align.f} to keep the chiral enhanced part \eqref{cR:lambda.f} vanishing, at the cost of devising a mass dependent condition. This vanishing could even be extended to the whole dipole coefficient \eqref{C.dipole}. 
We will keep the simple orthogonality condition \eqref{align.f}, and use the mass difference $\Delta M_N$ as a quantifier of the degree of tuning.

\begin{figure}[h]
\includegraphics[scale=0.5]{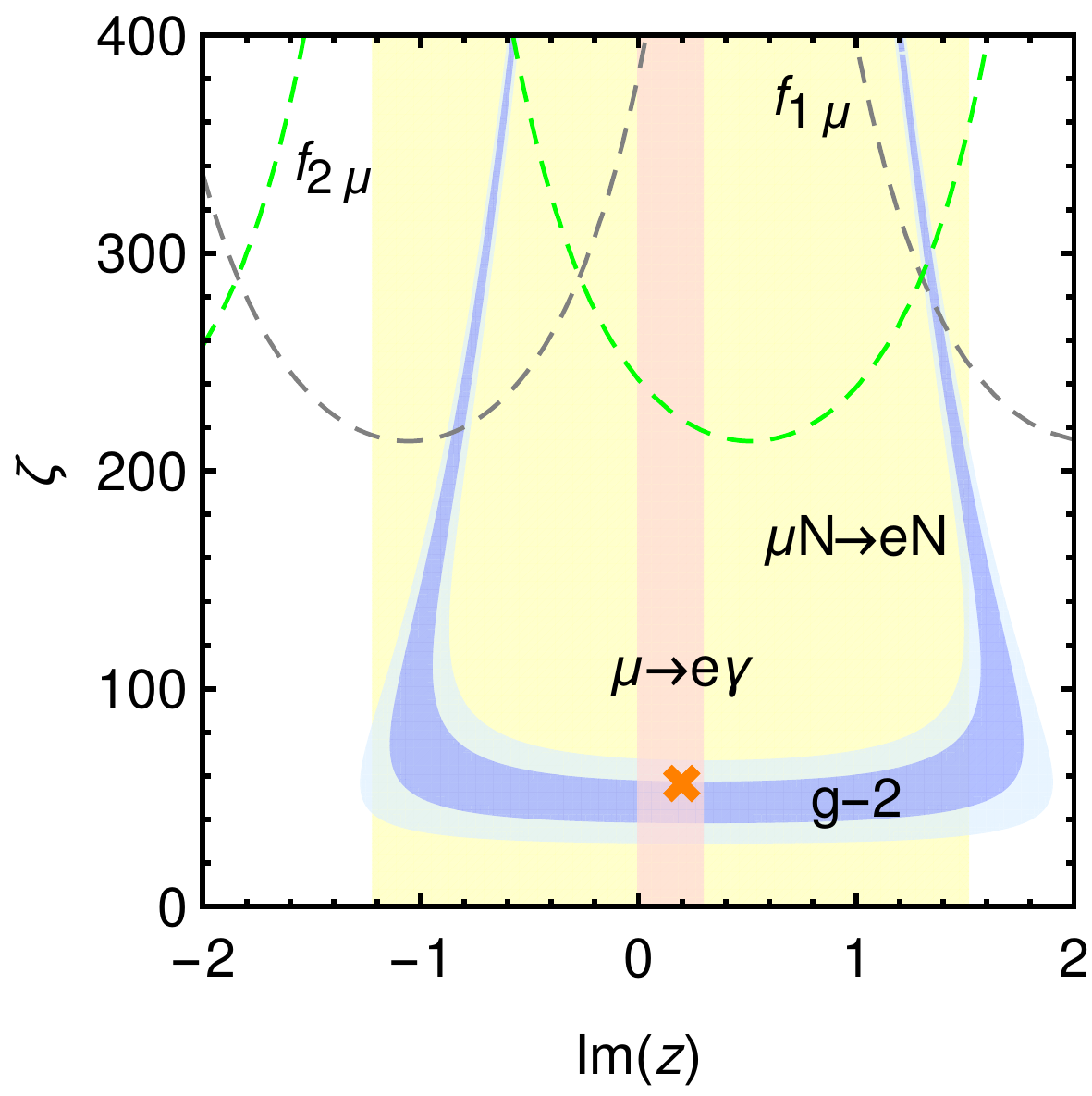}
\includegraphics[scale=0.5]{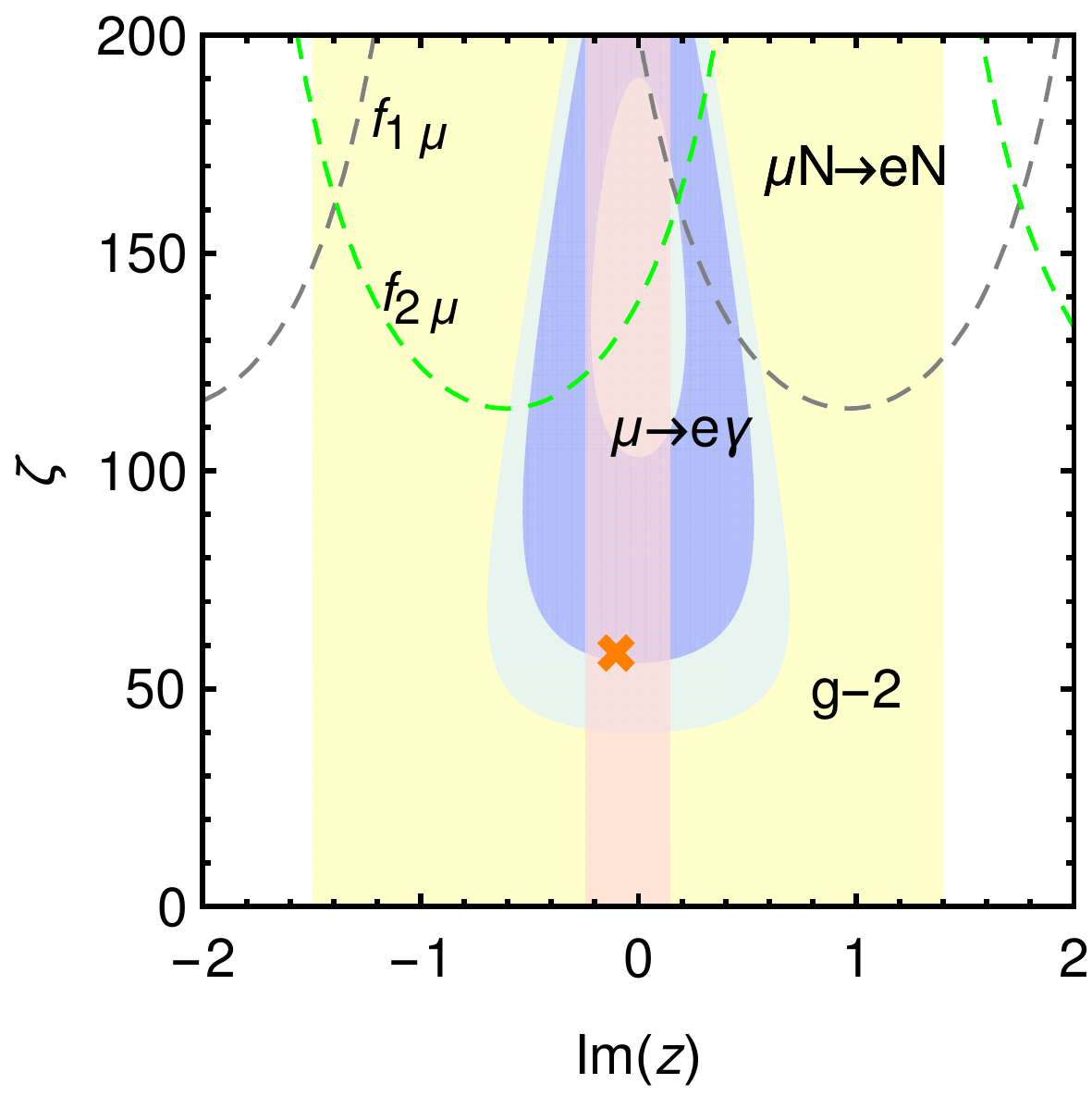}
\caption{\label{fig:z:bounds}%
Allowed regions in the plane $\im(z)\times\zeta$ for NO (left) and IO (right).
Scalar masses and $s_\gamma$ are fixed as \eqref{BM:NO:1} while $M_N=1\,\unit{TeV}$ and $v_{1}=x\times 10^{-3}\;\unit{GeV}$ with $x=1$ ($x=2$) for NO (IO). The region above the gray (green) dashed lines are excluded by perturbativity on the couplings $f_{1\mu}$ ($f_{2\mu})$. The orange region is allowed by the present constraint on $\mu N \to e N$, while the light red region is related to $\mu\rightarrow e\gamma$. The dark (light) blue region show the 1(2)-$\sigma$ region allowed by $(g-2)_\mu$. The crosses denote the benchmark points in \eqref{BM:NO:2}.
The mixing angles of the PMNS are fixed in best-fit of \cite{valle.fit} whereas $\delta=218^\circ$.
}
\end{figure}

With the orthogonality condition, the couplings $f_{i\alpha}$ are completely determined by free variables related to neutrinos masses ($\im(z)$, $M_{1}=M_{2}=M_N$), the vev $v_{1}$, and the scaling parameter $\zeta$. 
Note that the real part of $z$ is not physical for $M_1=M_2$.
Thus, for fixed values of $M_N$ and $v_{1}$, we can impose perturbativity bounds $|f_{i\alpha}|<4\pi$ as a function of $\im z$ and $\zeta$. If the masses of the charged scalars and their mixing is also fixed, we can obtain further bounds by requiring compatibility with $(g-2)_\mu$ and $\mu\rightarrow e\gamma$. 
For both NO and IO, we choose
\eq{
\label{BM:NO:1}
M_{S_{1}}=350\;\unit{GeV}\,,\quad M_{S_{2}}=450\;\unit{GeV}\,,\quad s_{\gamma}=0.1\,.
}
Figure~\ref{fig:z:bounds} shows allowed regions in the plane $\im(z){-}\zeta$ for $(g-2)_\mu$ (blue), $\mu\to e\gamma$ (pink) and $\mu e$ conversion in Au (yellow). Perturbativity for $f_{i\alpha}$ (dashed curves) is also shown.
The left (right) figure is for NO (IO) for which $M_N=1\,\unit{TeV}$ and $v_{1}=10^{-3}\;\unit{GeV}$ ($v_{1}=2\times 10^{-3}\;\unit{GeV}$).
The mixing angles of the PMNS matrix are fixed in the best-fit of \cite{valle.fit} while we choose $\delta=218^\circ$.
Using a different value of $\delta$ leads to different shapes and regions for this plot and others that follow but the overall possibility of explaining $(g-2)_\mu$ and avoiding CLFV do not change significantly.
The change is larger for NO. For IO the variation is not significant.

As illustrated in the plot, the strongest constraint on $\im(z)$ comes from $\mu\rightarrow e\gamma$. 
There is no visible constraint on $\zeta$ from $\mu\rightarrow e\gamma$ because the chiral enhanced term was chosen to vanish.
The constraint on $\zeta$ will mainly come from $(g-2)_\mu$, whose dominant chiral enhanced contribution depends linearly on this variable.
In turn, perturbativity bounds on $f_{ij}$ impose upper bounds on $\zeta$. 
For each NO and IO, we see that the following benchmark points account for $(g-2)_\mu$ still evading the current CLFV constraints:
\eqali{
\label{BM:NO:2}
\text{(BM-NO)}&\quad
M_N=1\,\unit{TeV}\,,\quad
v_1=10^{-3}\,\unit{GeV}\,,\quad
z=0.2i\,,\quad
\zeta=60\,;
\cr
\text{(BM-IO)}&\quad
M_N=1\,\unit{TeV}\,,\quad
v_1=2\times 10^{-3}\,\unit{GeV}\,,\quad
z=-0.1i\,,\quad
\zeta=60\,.
}
Only $v_1$ and $z$ differ in the two points. The rest of the model parameters are fixed as \eqref{BM:NO:1}.

\begin{figure}[h]
\includegraphics[scale=0.5]{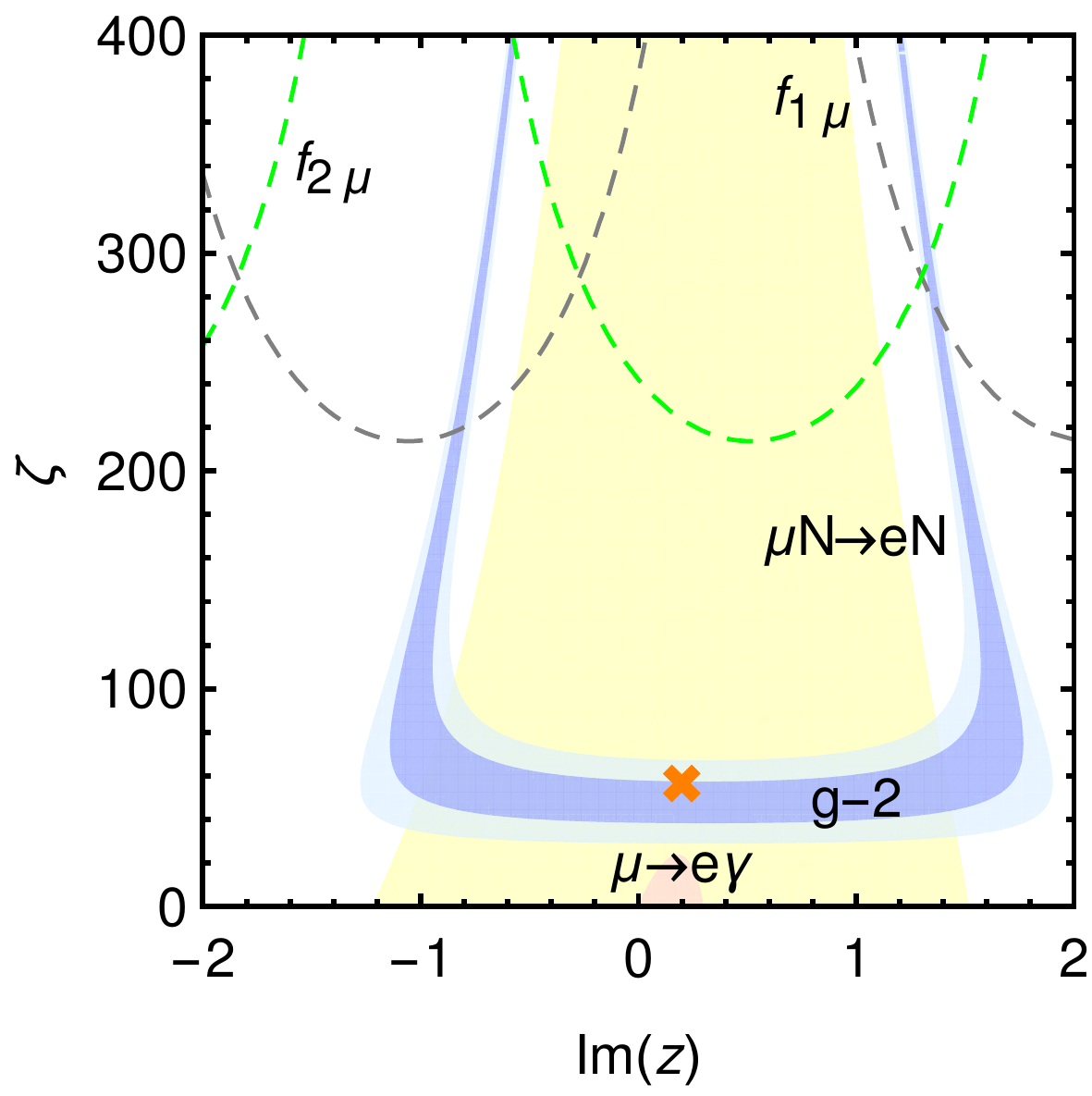}
\includegraphics[scale=0.5]{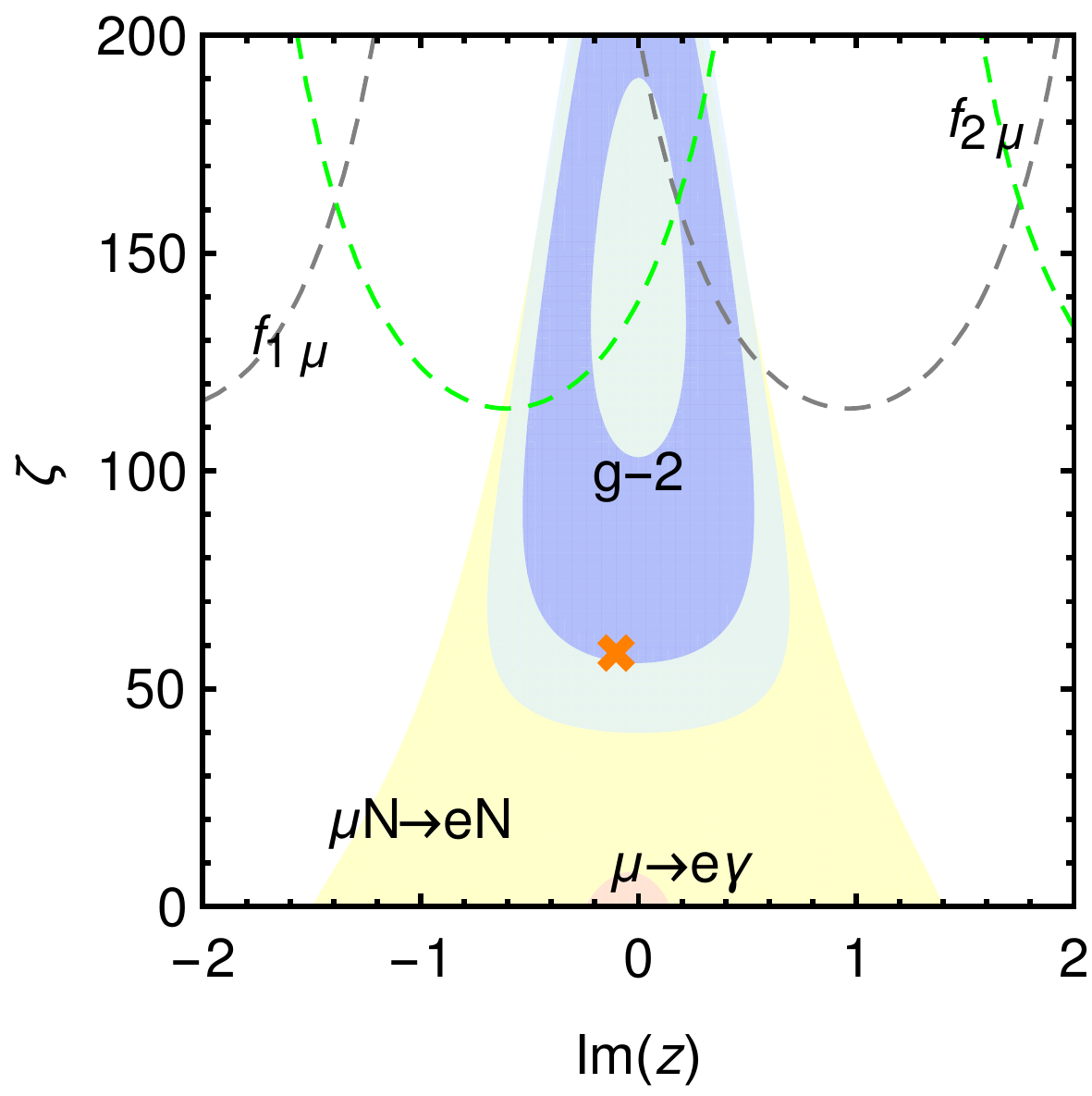}
\caption{\label{fig:z:bounds:nondege}%
The same as Fig.\,\ref{fig:z:bounds}, but for non-degenerate neutrino masses: $M_{2}=1.0001\,M_{1}$. 
Left: NO. Right: IO.
}
\end{figure}

For comparison and to assess the degree of tuning, we show in Fig.\,\ref{fig:z:bounds:nondege} the allowed regions analogous to Fig.\,\ref{fig:z:bounds}, with the same parameters, except for the lifting of the mass degeneracy of RHNs to $M_{2}=1.0001\,M_{1}$.
We can see that we lose compatibiliy between $(g-2)_\mu$ at 1$\sigma$ and $\mu\to e\gamma$ as the chiral enhanced term of the latter will not be completely canceled.
The region of compatibility for $(g-2)_\mu$ is practically unchanged.
We fix $\re(z)=0$ but we have checked that variation of $\re(z)$ may lead at most to a difference of a factor two.  

\begin{figure}[h]
\includegraphics[scale=1.0]{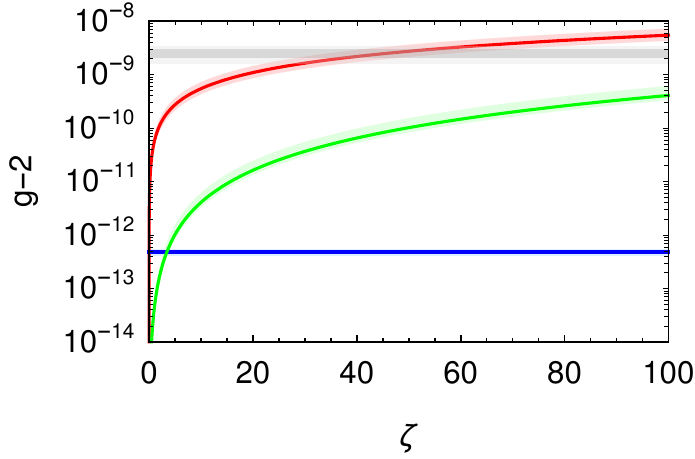}
\includegraphics[scale=1.0]{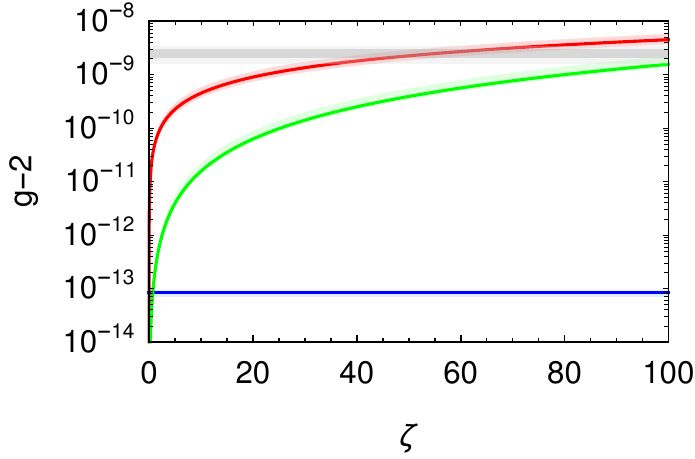}
\caption{The absolute value of the contributions for $(g{-}2)_\mu$ as defined in eq.\;\eqref{C.dipole} as a function of $\zeta$, for the benchmarks defined in \eqref{BM:NO:1} and \eqref{BM:NO:2} (left is NO and right is IO). The red line stands for the dominant chiral enhanced term \eqref{cR:lambda.f}. The blue line refers to the $\lambda^2$ term, \eqref{cR:lambda.lambda}, while the green line corresponds to the $f^2$ term, \eqref{cR:f.f}. Both the green and blue contributions are negative, while the red one is positive. The dark (light) gray band corresponds to the 1(2)-$\sigma$ region allowed at present for $(g{-}2)_\mu$. 
The curves assumes $M_1=M_2$ while the bands around them show the variation for $|M_{2}/M_{1}-1|\leq 0.4$.}
\label{fig:g2}
\end{figure}

\begin{figure}[h]
\includegraphics[scale=1.0]{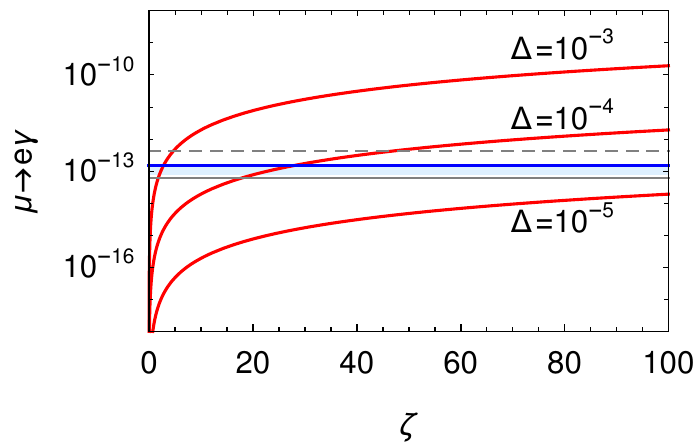}
\includegraphics[scale=1.0]{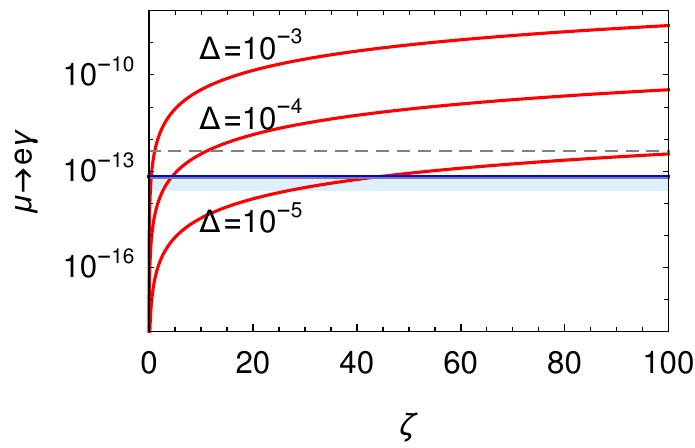}
\caption{The contributions for $\mu\to e\gamma$ as defined in eq.\;\eqref{C.dipole} as a function of $\zeta$, for the benchmarks defined in \eqref{BM:NO:1} and \eqref{BM:NO:2} (left is NO and right is IO). The red line stands for the chiral enhanced term \eqref{cR:lambda.f} while the blue line refers to the $\lambda^2$ term, cf.\,\eqref{cR:lambda.lambda}. The dashed (continuous) gray line corresponds to the present (future) limits. Here, $\Delta=M_{2}/M_{1}-1$. The band around the blue line shows how much this contribution varies if $\Delta$ varies within $|\Delta|\leq 0.4$.}
\label{fig:mueg}
\end{figure}

\begin{figure}[h]
\includegraphics[scale=1.0]{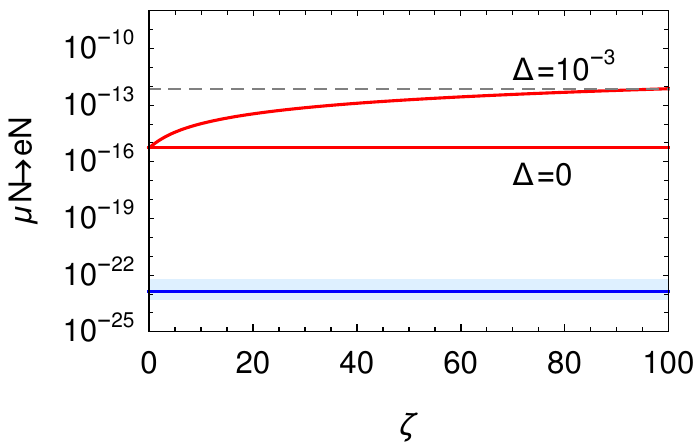}
\includegraphics[scale=1.0]{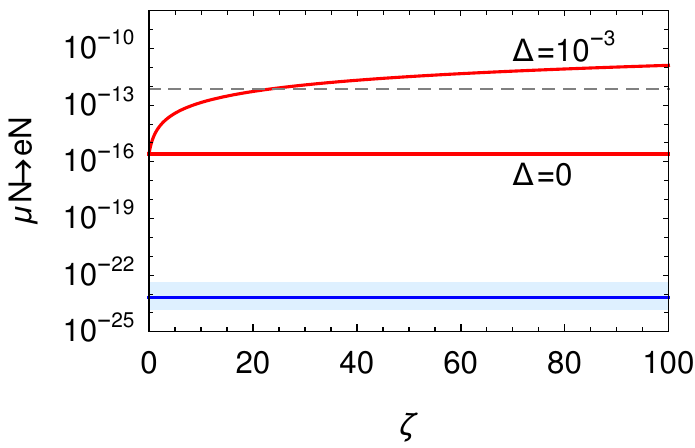}
\caption{The contributions for $\Br(\mu \rm{Au}{\to}e \rm{Au})$ as defined in eqs.\;\eqref{C.dipole} and \eqref{CND:nu} as a function of $\zeta$, for the benchmark defined in \eqref{BM:NO:1} and \eqref{BM:NO:2} (left is NO and right is IO). The red line stands for the dipole contribution \eqref{C.dipole}, which is dominant. Moreover, if the neutrino masses are chosen non-degenerate, it can violate the present bound, given by the dashed gray line. The blue line is related to the non-dipole contribution, \eqref{CND:nu}. Here, $\Delta=M_{2}/M_{1}-1$. The band around the blue line shows how much this contribution varies if $\Delta$ varies within $|\Delta|\leq 0.4$.}
\label{fig:conv}
\end{figure}

We illustrate in figures \ref{fig:g2}-\ref{fig:conv} the interplay of the distinct contributions to $g-2$, $\mu\rightarrow e\gamma$, and $\mu e$ conversion in Au respectively, using the benchmarks defined in \eqref{BM:NO:1} and \eqref{BM:NO:2}. In all the plots, we show the influence of the $\zeta$ parameter in the various contributions for each observable. Starting with the contributions to $g-2$, Fig.\,\ref{fig:g2}, we show that the chiral enhanced term given by \eqref{cR:lambda.f} is dominant, easily explaining the present anomaly. We also show in the figure that allowing for non-degenerate neutrino masses affects very little. Considering  $\mu\rightarrow e\gamma$, for non-degenerate masses the chiral contribution can easily surpass the bound on $\mu\rightarrow e\gamma$. For the case of degenerate masses, the present bound can be avoided for this benchmark, even though for the future limit this would not be the case. The important point to notice is that all the contributions can be suppressed by increasing $v_{1}$ since all of them are proportional to $v_{1}^{-2}$ under the orthogonality condition. However, since only the chiral contribution is proportional to $s_{2\gamma}$, we can still compensate by increasing the value of $s_{\gamma}$. This reasoning was applied when choosing the benchmark for IO, where $v_{1}$ is higher than the case for NO. At last, we comment on the present bounds on $\mu e$ conversion in Au, Fig.\,\ref{fig:conv}. We show that the non-dipole contribution given by \eqref{CND:nu} is negligible compared to the dipole contribution given by \eqref{C.dipole}. Moreover, as in the case of $\mu\rightarrow e\gamma$, if the neutrino masses are non-degenerate we can easily surpass the present bound.

To illustrate how the CLFV constraints are sensitive to the mass difference $M_2-M_1$ of the RHNs, we can choose for NO the scaling $v_1=10^{-3}\,\unit{GeV}\sqrt{M_{1}/\unit{TeV}}$ such that the Yukawas $\la{1}$ in \eqref{nu2HDM:lambda} has the overall scale fixed as
\eq{
(\la{1}{\la{1}}^\dag)_{11}=\frac{M_1}{v_1^2}\Big(
m_2|c_z|^2+m_3|s_z|^2
\Big)\sim 
0.05\times |s_z|^2\,,
}
and the benchmark \eqref{BM:NO:2} is attained for $M_1=1\,\unit{TeV}$.
For the benchmark value we can see that $\la{1}\sim O(0.1)$ and $f$ is larger by $\zeta$ if respecting the orthogonality condition.
For IO we choose twice the value for $v_1$ so that the benchmark \eqref{BM:NO:2} is also attained for $M_1=1\,\unit{TeV}$.

\begin{figure}[h]
\includegraphics[scale=0.5]{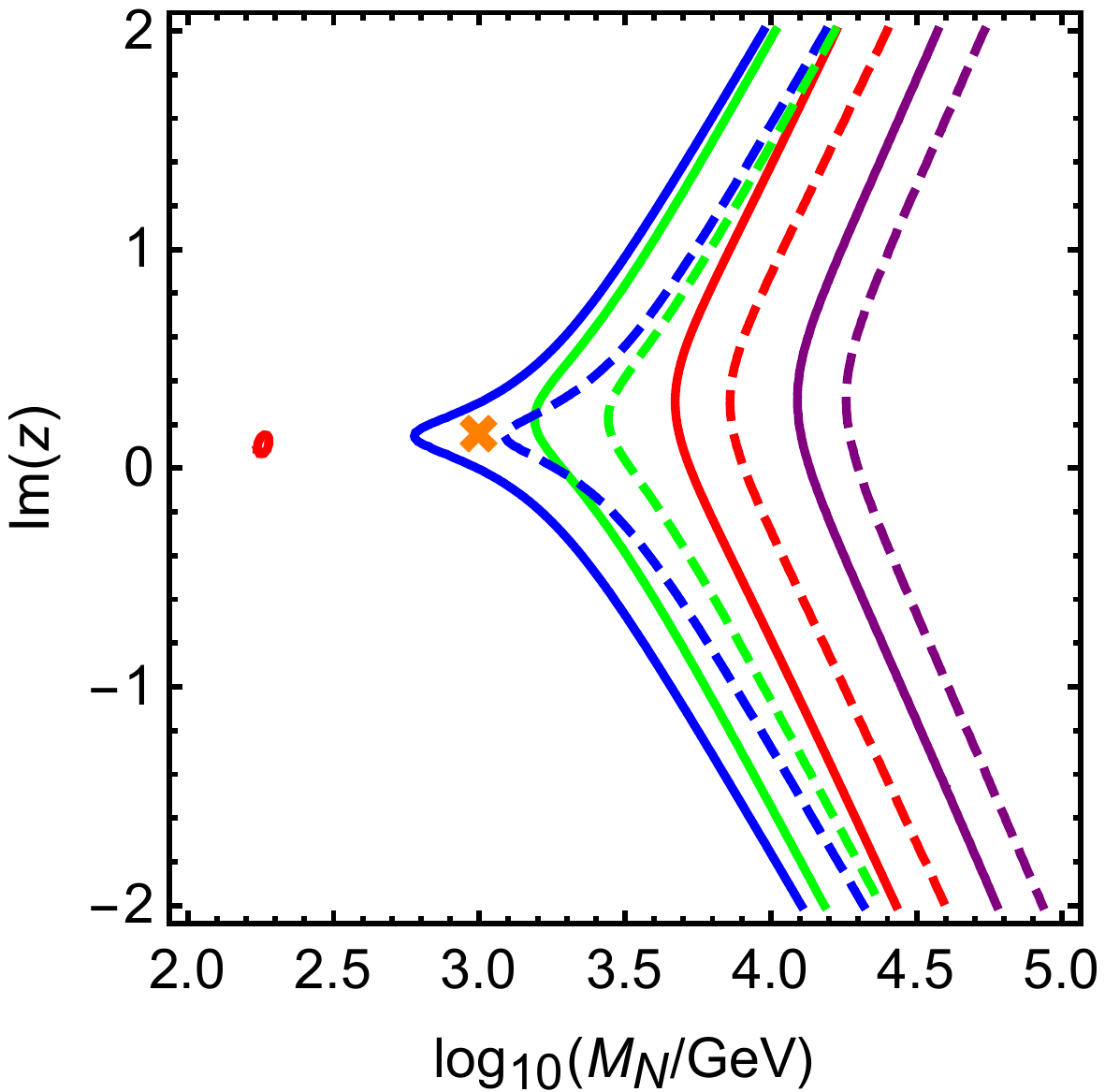}
\includegraphics[scale=0.5]{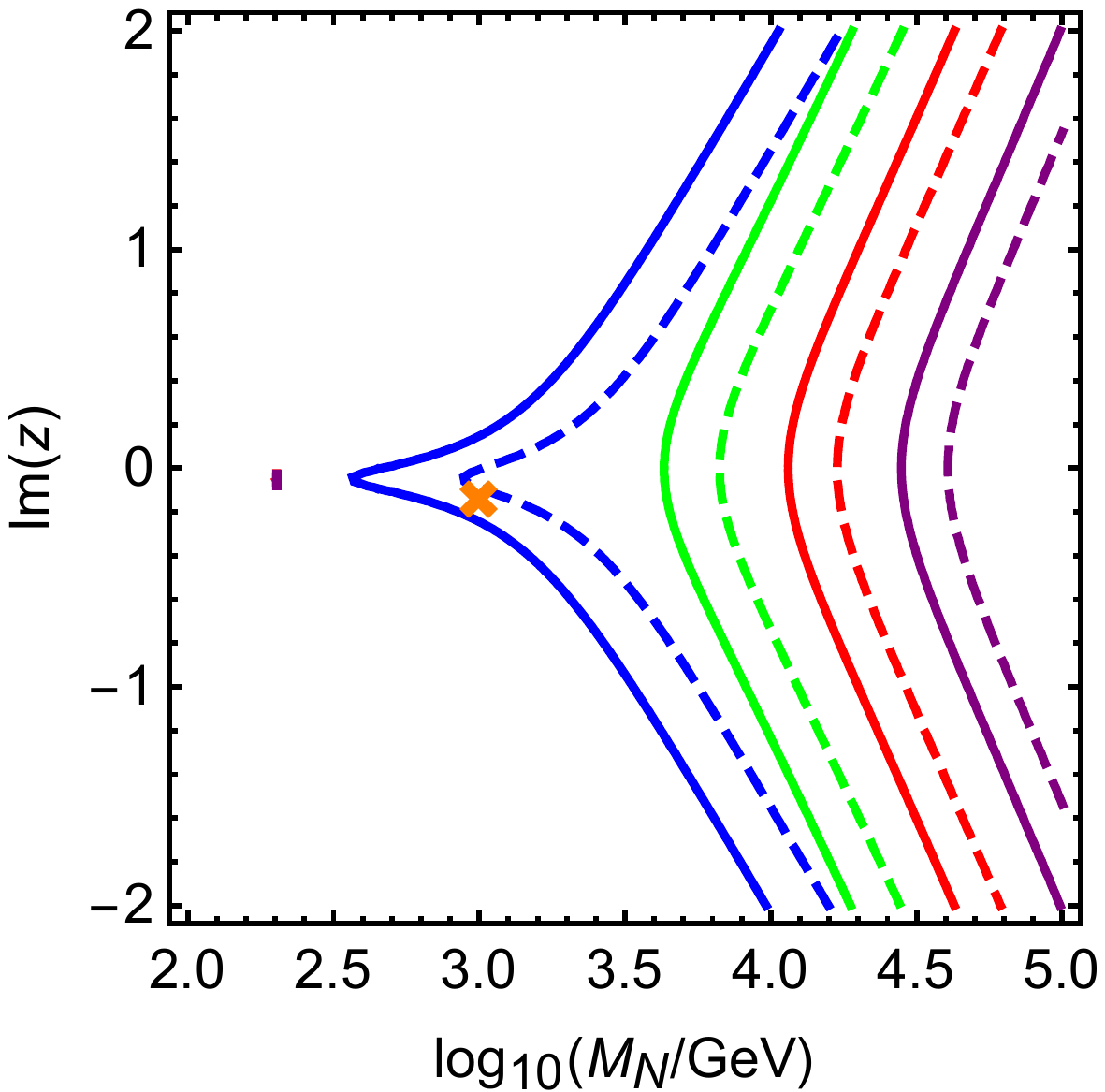}
\caption{\label{fig:muegamma:zMn}%
Contours for $\mu\to e\gamma$ for fixed $\zeta=60$, and varying $M_{2}/M_{1}-1=0, 10^{-4}, 10^{-3},10^{-2}$ (respectively in blue, green, red and purple).
The vev $v_1=x\times 10^{-3}\,\unit{GeV}\sqrt{M_{1}/\unit{TeV}}$ scales with $\sqrt{M_{1}}$ so that the global scale for the Yukawa $\la{1}$ is fixed.
The continuous curves denote $\Br(\mu\to e\gamma)=4.2\times 10^{-13}$ (current) and the dashed ones denote $\Br(\mu\to e\gamma)=6\times 10^{-14}$ (future).
The mass is $M_N=M_{1}$ and $z$ is the parameter in \eqref{nu2HDM:lambda} for NO (similarly for IO) with $\re(z)=0$. 
Left: NO with $x=1$. Right: IO with $x=2$.
The cross denotes the benchmark points defined in \eqref{BM:NO:1} and \eqref{BM:NO:2} for NO and IO.
}
\end{figure}

We start with countour curves, shown in Fig.\,\ref{fig:muegamma:zMn}, for the current and future limit for $\mu\to e\gamma$ as a function of $\im(z)$ and $M_{1}$ for different values of $\Delta=M_2/M_1-1$.
We clearly see that the curves move to the right as the mass difference increases showing that the constraints get stronger with the mass difference. 
The benchmark point, marked with a cross, is already excluded for $\Delta=10^{-4}$ for both NO and IO.
In the future, this benchmark will be excluded even for degenerate masses.
A curve similar to the blue curve can be also found in Ref.\,\cite{haba.yamada} where they consider the pure $\nu$-2HDM without the charged singlet and hence equivalent to our case turning off $f_{j\alpha}$. The shape of the curve is not exactly the same because their treatment of the Casas-Ibarra parametrization of the minimal case is not appropriate.
For certain curves, there are islands near $M_{1}\sim 10^{2.3}\,\unit{GeV}$ indicating a destructive interference between the chiral enhanced contribution and the $\lambda^2$ contribution in \eqref{C.dipole}.

\begin{figure}[h]
\includegraphics[scale=0.5]{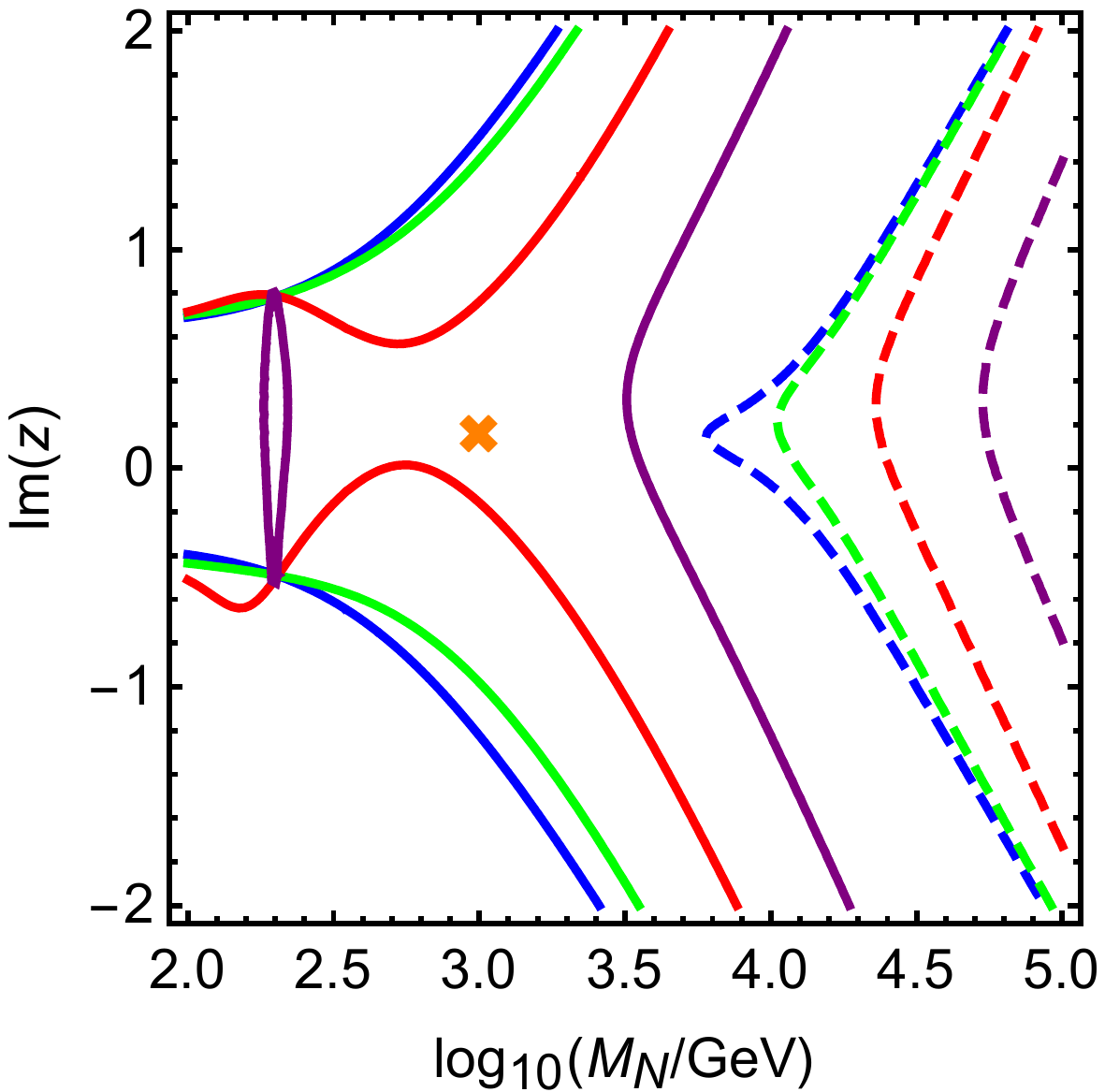}
\includegraphics[scale=0.5]{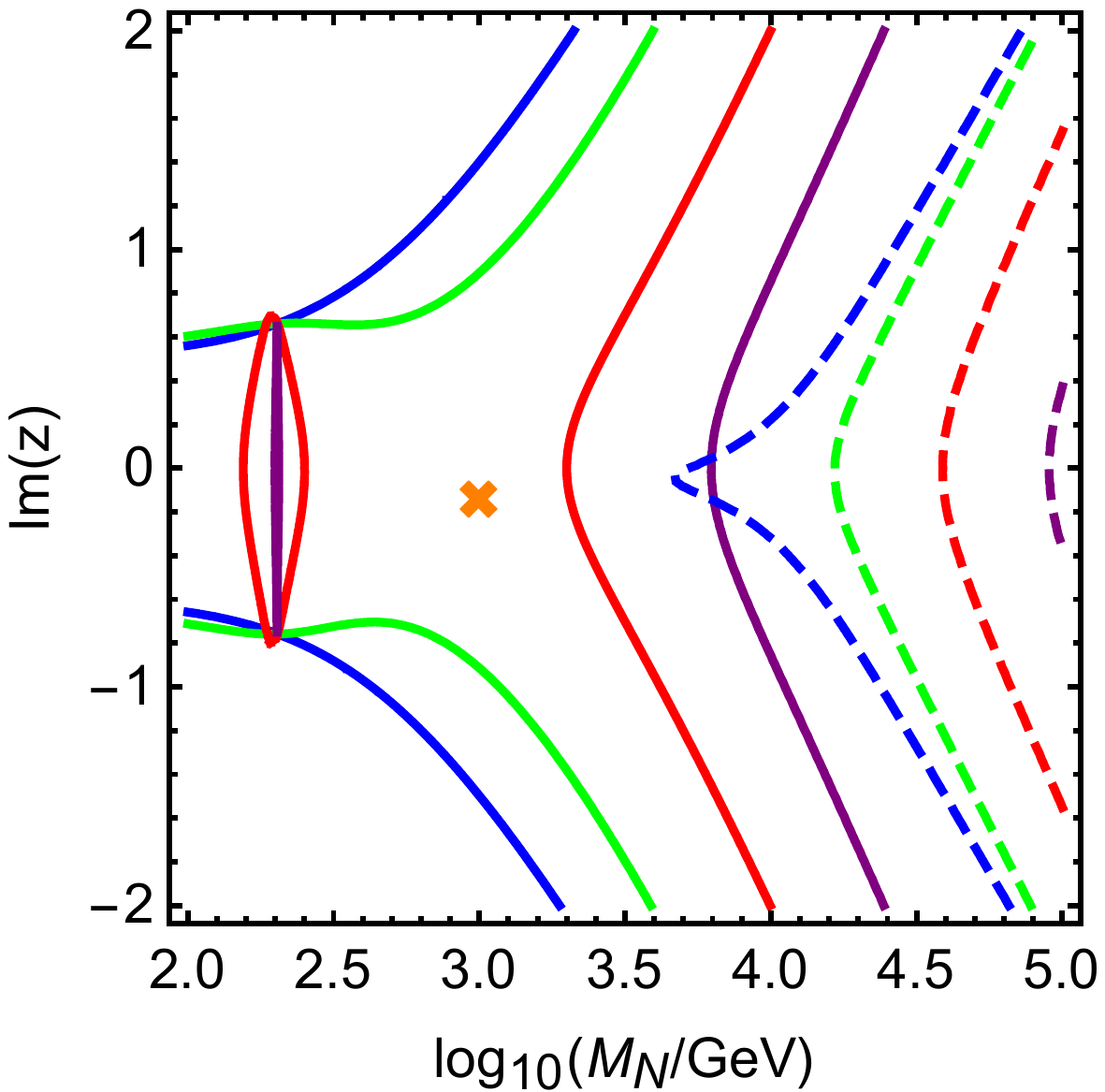}
\caption{\label{fig:mueconv:zMn}%
Contours for $\Br(\mu N{\to}e N)$ for fixed $\zeta=60$, and varying $M_{2}/M_{1}-1=0, 10^{-4}, 10^{-3},10^{-2}$ (respectively in blue, green, red and purple).
The vev $v_1=x\times 10^{-3}\,\unit{GeV}\sqrt{M_{1}/\unit{TeV}}$ scales with $\sqrt{M_{1}}$ so that the global scale for the Yukawa $\la{1}$ is fixed.
The continuous curves denote $N={\rm Au}$ (current) and the dashed denote $N={\rm Ti}$ (future).
The mass is $M_N=M_{1}$ and $z$ is the parameter in \eqref{nu2HDM:lambda} for NO (similarly for IO) with $\re(z)=0$. 
Left: NO with $x=1$. Right: IO with $x=2$.
The cross denotes the benchmark points defined in \eqref{BM:NO:1} and \eqref{BM:NO:2} for NO and IO.
}
\end{figure}

In Fig.\,\ref{fig:mueconv:zMn}, we show similar contours for $\mu e$ conversion in nuclei. We can see that the future constraints will be much stronger than the current ones.
Currently, a small mass difference is still allowed but in the future the benchmark will be easily excluded.

\begin{figure}[h]
\includegraphics[scale=0.95]{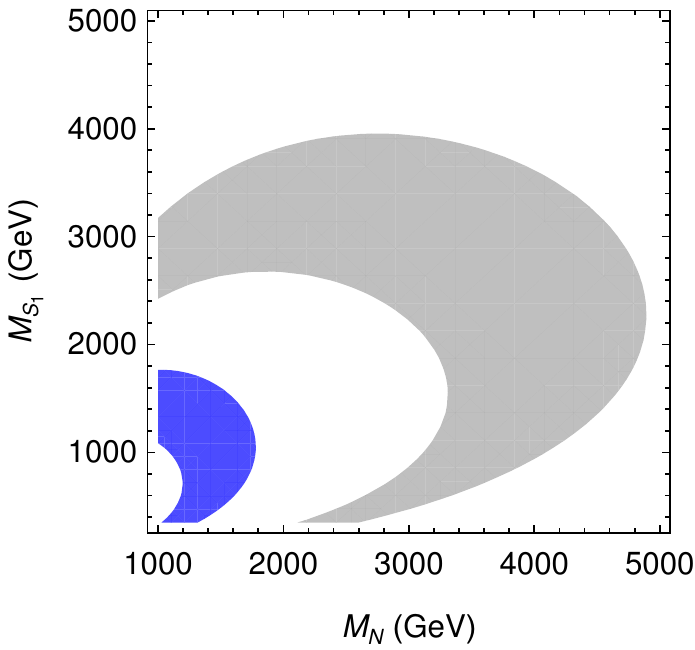}
\includegraphics[scale=0.95]{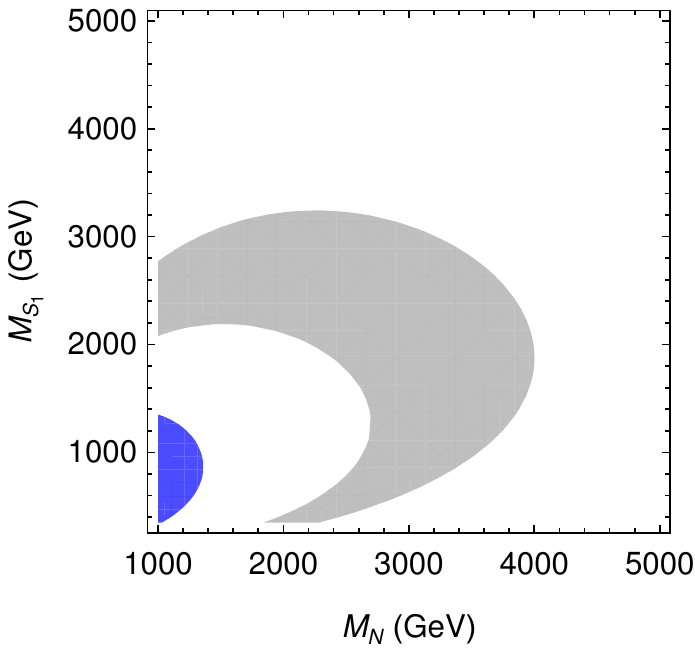}
\caption{\label{fig:1sigma:g-2}%
$1\sigma$ regions for $(g-2)_\mu$ in the $M_{S_1}$ and $M_N=M_1=M_2$ plane for fixed $\zeta=60$, $s_\gamma=0.1$ and two values of 
$M_{S_2}/M_{S_1}=450/350\approx 1.29$ (blue) and 
$M_{S_2}/M_{S_1}=2$ (gray).
The vev $v_1=x\times 10^{-3}\,\unit{GeV}\sqrt{M_{1}/\unit{TeV}}$ scales with $\sqrt{M_{1}}$ so that the global scale for the Yukawa $\la{1}$ is fixed.
Left: NO with $x=1$ and $z$ as in \eqref{BM:NO:2}. Right: IO with $x=2$ and $z$ as in \eqref{BM:NO:2}.
}
\end{figure}

Finally we briefly comment on the dependence of our results on the masses of the charged scalars.
The expression of the chiral enhanced contribution \eqref{cR:lambda.f} involves a cancellation between the contributions of the two charged scalars and it vanishes for degenerate masses.
So up to a certain point, increasing the mass difference leads to an increase in the contribution to $g-2$.
This information can be seen in Fig.\,\ref{fig:1sigma:g-2} where we show $1\sigma$ regions to satisfy the $(g-2)_\mu$ constraint in the $M_{S_1}$ and $M_N=M_1=M_2$ plane keeping the ratio fixed as $M_{S_2}/M_{S_1}=450/350\approx 1.29$ (blue) and $M_{S_2}/M_{S_1}=2$ (gray).
We can see that the larger mass ratio allows a larger compatibility region with larger masses for the scalars and the RHNs.
The benchmark points defined in \eqref{BM:NO:1} and \eqref{BM:NO:2} correspond to the origin of the plot in the corner of the blue region.
Following this point to the right inside the blue region would still allow the compatibility to $(g-2)_\mu$ but with decreasing contribution to CLFV.
We should remark that the mass difference of the charged scalars cannot be arbitrarily large as this would require an increasingly large $\mu_{\varphi}$. 
The benchmark we have chosen, given by (\ref{BM:NO:1}), is conservative in this sense, since it assumes $\mu_{\varphi} \sim -30\; \unit{GeV}$. For higher values, for instance, $-\mu_{\varphi} \sim \unit{TeV}$, it will still be possible to explain $(g-2)_\mu$ with masses of the charged scalars and the RHNs of $\mathcal{O}(\unit{TeV})$ given all other parameters fixed as before.
For even higher values of $-\mu_{\varphi} \sim 20\;\unit{TeV}$, we can avoid future constraints on $\mu e$ conversion in nuclei and still satisfy $(g-2)_\mu$.

Before ending this section, let us briefly discuss the collider phenomenology.
Given new particles with masses at the TeV scale, many BSM states can be searched for directly. Since the type IB seesaw model cannot account for $(g-2)_\mu$, 
we consider only the neutrinophilic version in eq.\,\eqref{lag:nuphilic} in which the neutrinophilic Higgs doublet $\Phi_1\sim H_\nu$ essentially contains only BSM states and their interactions with quarks are suppressed by the large $\tan\beta$, just as in the type I 2HDM. The interaction with quarks of the charged scalars within the singlet $\varphi^+$ are analogously suppressed and occurs only through mixing. 
Therefore, the dominant production mechanism at LHC for the charged scalars present in the extended neutrinophilic model will be through pair production $q\bar{q}\to \gamma^*(Z^*)\to S^-S^+$\,\cite{charged.higgs}, while the production of the gauge singlets $N_R$ will be highly suppressed\,\cite{deppisch} by the light-heavy neutrino mixing $|V_{\ell N}|\sim v_1\lambda^{(1)}/M_N\sim 10^{-6}$ for $M_N\sim \unit{TeV}$ and $v_1\sim \unit{MeV}$.
Therefore, we will not consider the decay of $N_R$.

The decay mode of the charged scalars $S_i^+$ depends on their masses relative to the masses of the RHNs.
If $M_{S_i}<M_{N_j}$, assumed in our benchmarks, and not so large $\tan\beta=v_2/v_1$, the decay of $S_i$ occurs dominantly 
through the Yukawa coupling to $\Phi_2$, which contains a tiny admixture of $S_i^+$. The dominant width to $tb$ quarks is roughly
\eq{
\Gamma(S_i^+\to tb)\sim \frac{y_t^2}{\tan^2\beta}\frac{M_{H^\pm}}{16\pi}\,,
}
where we ignore the suppression by $c_\gamma^2$ or $s_\gamma^2$.
For large $\tan\beta$, the dominant decay is through Yukawa couplings $\lambda^{(1)}$ and $f$ suppressed by the small mixing between the light and heavy neutrinos with width
\eq{
\label{S->lnu}
\Gamma(S_i^+\to \ell_\alpha^+\nu_{j})\approx\frac{M_{S_i}}{16\pi}\left[|\Gamma^H_{j\alpha}a_i|^2+|\Gamma^\varphi_{j\alpha}b_i|^2\right]\,,
}
where $\Gamma^H=v_1V^\tp {\la{1}}^\tp M_R^{-1}\la{1}=\hat{M}_\nu V^\dag/v_1$, $\Gamma^\varphi=v_1V^\dag{\la{1}}^\dag M_R^{-1}f^*$, $(a_1,a_2)=(c_\gamma,-s_\gamma)$, 
$(b_1,b_2)=(s_\gamma,c_\gamma)$.
The term with $\Gamma^H$ is already present in the $\nu$-2HDM without the charged singlet\,\cite{nu2HDM:pheno}.
As we need $f_{1\mu},f_{2\mu}\sim O(1)$ to solve $(g-2)_\mu$, the contribution from $\Gamma^\varphi$ is larger.
From a rough estimate, the decay \eqref{S->lnu} start to dominate for $\tan\beta\gtrsim 10^6$.
In contrast, if $M_{S_i}>M_{N_j}$, the charged scalars decay rapidly through the Yukawa couplings $\la{1}$ and $f$ into charged leptons and $N_R$. The latter will be long-lived and decay to $W^+\ell_\alpha$ or $Z\nu_j$ with rates suppressed by the light-heavy neutrino mixing.

To conclude this section, we have seen that the solution for $(g-2)_\mu$ within our model requires a sensitive balance of parameters to evade the strong bounds coming from CLFV processes.
In general, increasing $s_\gamma$, increasing $f_{j\alpha}$ or decreasing $v_1$ will increase the chiral enhanced contribution which dominates in $(g-2)_\mu$ but also in the CLFV processes if not properly suppressed.

\section{Summary}
\label{sec:summary}

In order to connect the mechanism of neutrino mass generation with the $(g-2)_\mu$ anomaly, we proposed to add a single charged singlet to the neutrinophilic 2HDM and a variant which implements a low scale seesaw by attributing part of the smallness of neutrino masses to a small vev.
We studied two models: (a) the $\nu$-2HDM version and (b) the type IB seesaw version, both for the minimal case of two righthanded neutrinos.
A chiral enhanced contribution to $(g-2)_\mu$ is generated through exchange of charged scalars and righthanded neutrinos, the latter also participating in the neutrino mass generation.
As family lepton number breaking is also brought to low scale, the chiral enhanced contribution generically leads to large rates for CLFV processes.
We find that the type IB seesaw implementation does not have enough freedom to circumvent the constraints from CLFV while solving the $(g-2)_\mu$ anomaly.
The $\nu$-2HDM version, on the other hand, has enough freedom to allow some special cases where $(g-2)_\mu$ anomaly can be solved and yet avoiding the stringent CLFV processes.
Even in these special cases, the region of compatibility between $(g-2)_\mu$ and current CLFV is very restricted.
One region for some choices of parameters is given in Fig.\,\ref{fig:z:bounds}.
It is clear the our solutions are not restricted to the minimal case of two RHNs as this limit can be mimicked in the presence of three righthanded neutrinos and more regions may open up.
In the future, experiments of $\mu e$ conversion in different nuclei are expected to drastically improve the limits and this kind of solution to the $(g-2)_\mu$ anomaly will be put to test.

\acknowledgments

A.C.\ acknowledges support from National Council for Scientific and Technological Development – CNPq through projects 166523\slash2020-8 and 201013\slash2022-3.
G.D.C.\ acknowledges financial support by the Coordenação de Aperfeiçoamento de Pessoal de Nível Superior - Brasil (CAPES) - Finance Code 001.
C.C.N.\ acknowledges partial support by Brazilian Fapesp, grant 2014/19164-6, and
CNPq, grant 312866/2022-4.

\appendix
\section{Wilson coefficients}
\label{ap:wilson}

Here we describe briefly how we obtain the Wilson coefficients of the effective photonic operators \eqref{L.eff:photon}.
The calculations are based on appendices \ref{ap:integrals} and \ref{ap:operators}.

\subsection{$\nu$-2HDM model}

For the dipole part, the contribution \eqref{cR:lambda.lambda} comes from the $LL$ part of the penguin diagram leading to \eqref{Imu:LL} with the following couplings multiplying \eqref{Csigma:LL}:
\eqali{
\label{couplings:LL}
N_k-S_1^+:&\quad -e{\la{1}_{\beta k}}^\dag\frac{v_2c_\gamma}{v}\la{1}_{k\alpha}\frac{v_2c_\gamma}{v}\,,
\cr
N_k-S_2^+:&\quad -e{\la{1}_{\beta k}}^\dag\frac{v_2(-s_\gamma)}{v}\la{1}_{k\alpha}\frac{v_2(-s_\gamma)}{v}\,.
}
The masses should be attributed accordingly.
The contribution \eqref{cR:f.f} comes from the $RR$ part of the penguin diagram leading to \eqref{Imu:LL}, $L$\,--\,$R$ exchanged, with the following couplings multiplying \eqref{Csigma:RR}:
\eqali{
\label{couplings:RR}
N_k-S_1^+:&\quad -ef^\tp_{\beta k}s_\gamma f^*_{k\alpha}s_\gamma\,,
\cr
N_k-S_2^+:&\quad -ef^\tp_{\beta k}c_\gamma f^*_{k\alpha}c_\gamma\,.
}
The chirally enhanced contribution \eqref{cR:lambda.f} comes from $LR$ part of the penguin diagram leading to \eqref{Imu:LR}, with the following couplings multiplying \eqref{Csigma:LR}:
\eqali{
N_k-S_1^+:&\quad -e {\la{1}_{\beta k}}^\dag\frac{v_2c_\gamma}{v} f^*_{k\alpha}s_\gamma\,,
\cr
N_k-S_2^+:&\quad -e {\la{1}_{\beta k}}^\dag\frac{v_2(-s_\gamma)}{v} f^*_{k\alpha}c_\gamma\,.
}

For the non-dipole part, there are only chirality preserving contributions.
The integral $I^{\mu}_{LL}$ \eqref{Imu:LL} and the coefficient \eqref{CND:LL} should be multiplied by \eqref{couplings:LL} while the analogous $I^{\mu}_{RR}$ should be multiplied by \eqref{couplings:RR},
resulting in the coefficients \eqref{CND:nu}.

\subsection{Seesaw type IB}

The contribution \eqref{cR:lambda.lambda:Ib} comes from the $LL$ part \eqref{Imu:LL} with the following couplings multiplying \eqref{Csigma:LL}:
\eqali{
\label{couplings:LL:Ib}
N-S_1^+:&\quad -e\lambda^\dag_{\beta 1}\frac{v_2c_\gamma}{v}\lambda_{1\alpha}\frac{v_2c_\gamma}{v}\,,
\cr
N-S_2^+:&\quad -e\lambda^\dag_{\beta 1}\frac{v_2(-s_\gamma)}{v}\lambda_{1\alpha}\frac{v_2(-s_\gamma)}{v}\,.
}
The contribution \eqref{cR:f.f:Ib} comes from the $RR$ part of \eqref{Imu:LL}, $L$\,--\,$R$ exchanged, with the following couplings multiplying \eqref{Csigma:RR}:
\eqali{
\label{couplings:RR:Ib}
N-S_1^+:&\quad -ef^\tp_{\beta 2}s_\gamma f^*_{2\alpha}s_\gamma\,,
\cr
N-S_2^+:&\quad -ef^\tp_{\beta 2}c_\gamma f^*_{2\alpha}c_\gamma\,.
}
The chirally enhanced contribution \eqref{cR:lambda.f:Ib} comes from $LR$ part of the penguin diagram leading to \eqref{Imu:LR}, with the following couplings multiplying \eqref{Csigma:LR}:
\eqali{
N-S_1^+:&\quad -e \lambda^\dag_{\beta 1}\frac{v_2}{v}c_\gamma f^*_{2\alpha}s_\gamma\,,
\cr
N-S_2^+:&\quad -e \lambda^\dag_{\beta 1}\frac{v_2}{v}(-s_\gamma) f^*_{2\alpha}c_\gamma\,.
}
Other combinations are forbidden by the pseudo-Dirac nature of $N$.

For the non-dipole part, there are only chirality preserving contributions.
The integral $I^{\mu}_{LL}$ \eqref{Imu:LL} and the coefficient \eqref{CND:LL} should be multiplied by \eqref{couplings:LL} while the analogous $I^{\mu}_{RR}$ should be multiplied by \eqref{couplings:RR},
resulting in the coefficients \eqref{CND:Ib}.

\section{Loop integrals}
\label{ap:integrals}

Our calculations for loop integrals and operators are similar to Refs.\,\cite{bilenky,raidal.santamaria}.
The following loop integrals come from self-energy diagrams and penguin diagrams in Fig.\,\ref{fig:dipole.self}:
\subeqali[I.self]{
\label{I:LL}
iI_{LL}&=\int \frac{d^4k}{(2\pi)^4}\frac{R\slashed{k}}{(k^2-M_N^2)[(k-p)^2-\Mphi^2]}
\cr
&=\frac{iR}{(4\pi)^2}\,\frac{\slashed{p}}{2}\left\{\frac{1}{\epsilon}+\log\frac{\bar{\mu}^2}{\Mphi^2}
+h_S(x)+\frac{p^2}{\Mphi^2}8\tf_S(x)\right\}\,,
\\
\label{I:LR}
iI_{LR}&=\int \frac{d^4k}{(2\pi)^4}\frac{R M_N}{(k^2-M_N^2)[(k-p)^2-\Mphi^2]}
\cr
&=\frac{iR}{(4\pi)^2}M_N\left\{\frac{1}{\epsilon}+\log\frac{\bar{\mu}^2}{\Mphi^2}
+\frac{1-x+x\log x}{1-x}+\frac{p^2}{\Mphi^2}2f_S(x)\right\}\,,
}
\subeqali[I.dipole]{
\label{Imu:LL}
iI^\mu_{LL}&=\int \frac{d^4k}{(2\pi)^4}\frac{R\slashed{k}(p_1+p_2-2k)^\mu}{(k^2-M_N^2)[(k-p_1)^2-\Mphi^2][(k-p_2)^2-\Mphi^2]}
\cr
&=-\frac{iR}{(4\pi)^2}\bigg\{
    \ums{2}\gamma^\mu\left[\frac{1}{\epsilon}+\log\frac{\bar{\mu}^2}{\Mphi^2}+h_S(x)\right]
    +\frac{1}{M_\varphi^2}(q^2\gamma^\mu-\slashed{q}q^\mu)\ums{6}G_S(x)
\cr&\qquad\qquad
    +\ \frac{1}{M_\varphi^2}[(\slashed{p_1}+\slashed{p_2})(p_1+p_2)^\mu+(p_1^2+p_2^2)\gamma^\mu]2 \tf_S(x)
\bigg\}\,,
\\
\label{Imu:LR}
iI^\mu_{LR}&=\int \frac{d^4k}{(2\pi)^4}\frac{R M_N(p_1+p_2-2k)^\mu}{(k^2-M_N^2)[(k-p_1)^2-\Mphi^2][(k-p_2)^2-\Mphi^2]}
\cr
&=-\frac{iR}{(4\pi)^2}\frac{M_N}{\Mphi^2}(p_1+p_2)^\mu 2f_S(x)
\,,
}
where $x=M_N^2/\Mphi^2$ and $q=p_2-p_1$. 
These expression should be supplied with couplings and enclosed by spinors $\bar{u}(p_2)$ and $u(p_1)$ to give the amplitudes.
Simple chirality exchange $L\leftrightarrow R$ leads to identical expressions with the projector exchanged.
We use dimensional regularization with $d=4-2\epsilon$ 
and retain only terms up to $\Mphi^{-2}$ or $M_N^{-2}$.
The additional loop function that appear is
\eq{
h_S(x)=\frac{1-4x+3x^2-2x^2\log x}{2(1-x)^2}\,.
}
The result for $I_{LL}$ and $I^\mu_{LL}$ match \cite{bilenky} for $x\to 0$ as $8\tf_S(0)=1/3$ and $G_S(0)=1/3$.

For the last integral in \eqref{Imu:LR}, we can use the Gordon-type identity where we can replace
\eq{
\label{gordon}
(p_1+p_2)^\mu\to -i\sigma^{\mu\nu}q_\nu+\slashed{p_2}\gamma^\mu+\gamma^\mu\slashed{p_1}\,.
}

\section{Operators}
\label{ap:operators}

The operators relevant to CLFV are the photon interactions written in \eqref{L.eff:photon}.
The operator in the first line is the dipole contribution whereas the ones in the second line are the non-dipole (ND) part.
These Wilson coefficients at 1-loop can be obtained by matching the full theory with the effective theory through appropriate 1-loop amplitudes.
The relevant amplitudes lead to the expressions in \eqref{I.self} and \eqref{I.dipole} which comes from the self-energy and dipole diagrams in Fig.\,\ref{fig:dipole.self}.
To obtain the full expressions, one only needs to add the coupling constants and adapt the masses.

Let us focus first on the chirality preserving contributions.
The coefficient of $\slashed{p}$ in \eqref{I:LL} and of the $\gamma^\mu$ part in \eqref{Imu:LL} give rise to the operator $\bar{\psi}_L\slashed{D}\psi_L$, $\psi$ being the collection of lepton fields, and it should be removed by wave function renormalization.
The coefficient of $\slashed{p}p^2$ in \eqref{I:LL} and the last term in the square brackets in \eqref{Imu:LL}
lead to the operator $\bar{\psi}_L(\slashed{D}D^2+D^2\slashed{D})\psi_L$\,\cite{bilenky}.
This operator can be replaced by $\bar{\psi}_L\slashed{D}^3\psi_L$ with additional dipole contributions.
The former operator does not lead to relevant physical phenomena.
The result is that the last term in $I^\mu_{LL}$ \eqref{Imu:LL} generates
\eq{
\label{Csigma:LL}
\delta C^{\sigma R}_{\beta\alpha}=\frac{(-1)}{(4\pi)^2}\frac{1}{2M^2_\varphi}2\tf_S(x)m_\alpha\,,
}
while an analogous $I^\mu_{RR}$ leads to 
\eq{
\label{Csigma:RR}
\delta C^{\sigma R}_{\beta\alpha}=m_\beta\frac{(-1)}{(4\pi)^2}\frac{1}{2M^2_\varphi}2\tf_S(x)\,.
}
These contributions are chirality flipping after the use of the equations of motion.
Finally, the contribution proportional to $G_S(x)$ in \eqref{Imu:LL} leads to the non-dipole term with
\eq{
\label{CND:LL}
C_{\beta\alpha}^{\text{ND-}L}=\frac{(-1)}{(4\pi)^2}\frac{1}{6M^2_\varphi}G_S(x)\,.
}
A similar term comes from $I^\mu_{RR}$.

Let us now turn to the chirality flipping contributions.
The coefficient of $M_N$ in \eqref{I:LR} should be removed by lepton mass renormalization.
The coefficient of $p^2M_N$ in \eqref{I:LR} and the coefficient of $\slashed{p_2}\gamma^\mu+\gamma^\mu\slashed{p_1}$ in \eqref{Imu:LR}, after replacement \eqref{gordon}, leads to the operator $\bar{\psi}_L\slashed{D}^2\psi_R$ which is not phenomenologically relevant.
The dipole contribution \eqref{gordon} in \eqref{Imu:LR} leads to
\eq{
\label{Csigma:LR}
\delta C^{\sigma R}_{\alpha\beta}=\frac{1}{(4\pi)^2}\frac{M_N}{M^2_\varphi}2f_S(x)\,.
}


\end{document}